\documentclass[10pt,journal]{IEEEtran}
\usepackage{amssymb}
\usepackage{amsmath}
\usepackage{cite}
\usepackage{url}
\usepackage{hyperref}
\usepackage{xcolor}
\usepackage{cite,graphicx}
\usepackage{fancyhdr}
\usepackage{cases}
\usepackage{mdwmath}
\usepackage{mdwtab}
\usepackage{cuted}
\usepackage{caption}
\usepackage[scr=rsfs]{mathalpha}
\usepackage{amsthm}
\usepackage{setspace}
\usepackage[linesnumbered,ruled,vlined]{algorithm2e}
\usepackage{subfig}
\usepackage{multirow}
\usepackage{makecell}
\usepackage{mathtools}

\graphicspath{{./src/img/}}

\newtheorem{remark}{Remark}
\newtheorem{theorem}{Theorem}

\newtheorem{lemma}{Lemma}

\newtheorem{corollary}{Corollary}

\allowdisplaybreaks
\makeatother
\def\BibTeX{{\rm B\kern-.05em{\sc i\kern-.025em b}\kern-.08em
		T\kern-.1667em\lower.7ex\hbox{E}\kern-.125emX}}
\expandafter\def\expandafter\normalsize\expandafter{%
	\normalsize%
	\setlength\abovedisplayskip{4pt}%
	\setlength\belowdisplayskip{4pt}%
	\setlength\abovedisplayshortskip{2pt}%
	\setlength\belowdisplayshortskip{2pt}%
}

\title{Cram\'er-Rao Bound Optimization for Near-Field Sensing with Continuous-Aperture Arrays}

\author{Hao~Jiang,~\IEEEmembership{Graduate Student Member,~IEEE}, Zhaolin Wang,~\IEEEmembership{Member,~IEEE}, \\ Yuanwei Liu,~\IEEEmembership{Fellow,~IEEE}, and Arumugam Nallanathan,~\IEEEmembership{Fellow,~IEEE}

\thanks{
Hao Jiang, Zhaolin Wang, and Arumugam Nallanathan are with the School of Electronic Engineering and Computer Science, Queen Mary University of London, London E1 4NS, U.K. (e-mail: \{hao.jiang; zhaolin.wang; a.nallanathan\}@qmul.ac.uk).

Yuanwei Liu is with the Department of Electrical and Electronic Engineering, The University of Hong Kong, Hong Kong (e-mail: yuanwei@hku.hk).}}

\begin{document}
\maketitle
\begin{abstract}
A Cramér-Rao bound (CRB) optimization framework for near-field sensing (NISE) with continuous-aperture arrays (CAPAs) is proposed.
In contrast to conventional spatially discrete arrays (SPDAs), CAPAs emit electromagnetic (EM) probing signals through continuous source currents for target sensing, thereby exploiting the full spatial degrees of freedom (DoFs). 
The maximum likelihood estimation (MLE) method for estimating target locations in the near-field region is developed. 
To evaluate the NISE performance with CAPAs, the CRB for estimating target locations is derived based on continuous transmit and receive array responses of CAPAs. 
Subsequently, a CRB minimization problem is formulated to optimize the continuous source current of CAPAs. 
This results in a non-convex, integral-based functional optimization problem. 
To address this challenge, the optimal structure of the source current is derived and proven to be spanned by a series of basis functions determined by the system geometry. 
To solve the CRB minimization problem, a low-complexity subspace manifold gradient descent (SMGD) method is proposed, leveraging the derived optimal structure of the source current.
Our simulation results validate the effectiveness of the proposed SMGD method and further demonstrate that i)~the proposed SMGD method can effectively solve the CRB minimization problem with reduced computational complexity, and 
ii)~CAPA achieves a tenfold improvement in sensing performance compared to its SPDA counterpart, due to full exploitation of spatial DoFs.
\end{abstract}

\begin{IEEEkeywords}
    Cramér-Rao bound optimization, continuous-aperture array (CAPA), near-field sensing (NISE).
\end{IEEEkeywords}
\section{Introduction}
\IEEEPARstart{R}APID evolution of communication technologies is steering the development of the next-generation telecommunication networks to support not only high-speed data transmission but also advanced sensing capabilities \cite{jiang2024terahertz}. 
The unprecedented performance gains offered by multiple-input and multiple-output (MIMO) in communication inspire its application to sensing.
Compared with conventional sensing setups, MIMO-based sensing offers a high spatial resolution while alleviating the reliance on formidable bandwidth requirements \cite{amico2022cramer, wang2024rethinking}.
Given these benefits, MIMO has become a pivotal technology for 6G sensing.

In contrast to the current 5G networks operating at the 3.5 GHz band, the upcoming 6G networks are expected to operate at high-frequency bands, such as mmWave bands and sub-terahertz bands \cite{r3_5}.
These bands can provide extensive spectrum availability for the communication functionality and enable high spatial resolution for the integrated sensing functionality \cite{amico2022cramer}.
However, despite the resultant high throughput on the high-frequency bands, the atmospheric-induced attenuation severely degrades systems' performance \cite{wang2028millimeter}.
To overcome transmit power dissipation, modern wireless networks require massive MIMO systems with hundreds or thousands of antennas at mmWave and THz bands to achieve higher beamforming, multiplexing, and diversity gains, as discussed in \cite{r3_4,r3_5}, thus driving an evolution from massive MIMO (mMIMO) to gigantic MIMO (gMIMO).
These antenna configurations are variants of spatially discrete antenna arrays (SDPAs), where the half-wavelength spacing between antennas is consistently fulfilled.
While SDPAs are easy to fabricate, two key limitations arise: i) storing and optimizing hundreds of parameters that scale with antenna numbers, leading to high computational cost and latency at practical base stations, and ii) spatial degrees-of-freedom (DoFs) confined by the discrete placement of antenna elements \cite{r3_6}.

Recently, holographic MIMO (HMIMO) has emerged as a promising technology \cite{r3_1}, reducing antenna spacing within a fixed array aperture. Unlike conventional SDPAs, HMIMO features a nearly continuous aperture, allowing the current distribution to be tailored for diverse functionalities \cite{an2023tut1, an2023tut2}, including communication and sensing \cite{huang2020holo, zhang2022holo}. The nearly continuous aperture provides enhanced spatial DoFs \cite{r3_1, gong2024holographic}, driving fundamental changes in channel modeling \cite{an2023tut2, gong2024near} and analytical methods \cite{wei2024electromagnet}. 
As a specific form of HMIMO, the continuous aperture array (CAPA) further advances this concept by transmitting signals over a truly continuous aperture, enabling full exploitation of the array’s spatial DoFs.
Different from SDPAs, CAPA resolves these challenges by densifying elements to a sub-wavelength scale, thus forming a continuous aperture \cite{r3_1,r3_7}. 
The source current in CAPA is modeled as a continuous function rather than discrete variables, enabling seamless phase control, reduced optimization complexity, and full exploitation of spatial DoFs as antenna density grows unbounded \cite{r3_3,r3_8}. 
Practically, CAPA can be realized using electrically-, optically-, or acoustically-driven technologies \cite{r3_10,r3_11,r3_12}.
As a newly emerging technology, CAPA has drawn great attention from the telecommunications community, bringing huge communication enhancement \cite{zhao2025continuous}.
However, the sensing capability of CAPA only receives limited attention.
On the one hand, due to the continuous nature of source distributions on CAPA, the analysis of CAPA-based sensing can be fundamentally different from its SDPA-based counterpart;
On the other hand, with enhanced design DoF, CAPA has the potential to strengthen the sensing functionality of wireless networks. Motivated by this consideration, this paper focuses on characterizing the sensing performance bound in monostatic sensing setups.

\subsection{Prior Works}
\subsubsection{Near-Field Sensing with SPDAs}
NISE with spatially discrete arrays (SPDAs) has become a ``hotspot" of research and has been extensively investigated \cite{huang1991near, zhang2018localization, wang2023near, korso2010conditional, Khamidullina2021condition, wang2024near, Giovannetti2024performance}.
Specifically, the authors of \cite{huang1991near} demonstrated that, unlike the far-field angle-only sensing, NISE enables joint sensing of targets' angle and distance by exploiting the curvature of the impinging spherical wavefront in the near-field regions.
Building on this unique polar-domain dependence, the authors of \cite{zhang2018localization} proposed a low-complexity multiple signal classification (MUSIC) algorithm to localize multiple sensing targets, where a two-dimensional (2D) MUSIC spectrum is searched.
In a separate study, the authors of \cite{wang2023near} developed a sensing-performance optimization algorithm for near-field integrated sensing and communication (ISAC) systems.
To quantify the sensing performance analytically, the Cramér–Rao bound (CRB) is widely adopted in NISE tasks as the lower bound for unbiased estimators.
Specifically, the authors of \cite{korso2010conditional} and \cite{Khamidullina2021condition} derived the conditional and unconditional CRBs for mono-static and bi-static NISE systems.
The above studies are within the scope of low-mobility or static sensing scenarios, where the sensing target is assumed to remain stationary for a relatively long time.
For the mobility NISE tasks, \cite{wang2024near} and \cite{wang2024rethinking} showed that it is possible to simultaneously estimate both radial and transverse velocities of a moving target by leveraging the Doppler frequencies.
On top of this observation, the authors of \cite{wang2024near} proposed a predictive beamforming scheme for near-field mobility networks, while \cite{Giovannetti2024performance} analyzed the CRBs for the radial and transverse velocity sensing in such scenarios.
 
\subsubsection{Near-Field Sensing with CAPAs}
Unlike the extensive research efforts for NISE in SPDAs, research on CAPA-based NISE is still in its early stages and primarily focuses on active positioning scenarios, where the sensing target must actively transmit a dedicated signal to the receiver.
Due to the electromagnetic (EM)-based channel modeling for CAPAs, the channel response (or received electric field) is a function of both distance and angle, allowing for joint angle and distance estimation in the near-field regions.
For instance, the authors of \cite{amico2022cramer} analyzed a CAPA-based positioning system, where the target (a Hertzian dipole) transmits probing signals actively to a holographic surface with a continuous aperture.
In light of the EM theory, CRB was derived for this setup, providing the lower bound for maximum likelihood estimation (MLE) of the target's position.
Considering both scalar and vector electric field scenarios, the authors of \cite{chen2024cramer} extended the analysis by deriving CRBs for a similar active positioning system in the near-field region.
This work demonstrated that centimeter-level sensing accuracy can be achieved in the near field of a receiving antenna array with a practical aperture size operating in mmWave or sub-THz bands.
Additionally, the authors of \cite{chen2024near} investigated the expected CRB (ECRB) and Ziv-Zakai bound (ZZB) for joint position and attitude estimation within the near-field region, considering a similar active sensing setup.
Beyond conventional SDPAs, CAPAs with continuous apertures enhance sensing performance by providing higher spatial resolutions. 
Moreover, by modeling an infinite number of antenna elements, CAPAs help characterize the fundamental sensing performance limits of MIMO systems. 
In addition, since CAPA optimization involves continuous functions rather than a large set of discrete variables, it significantly reduces memory requirements during the optimization process.

\subsection{Motivations and Contributions}
Despite the fruitful research endeavors in prior works, the majority of the existing literature on CAPA-based NISE falls in the category of active sensing, where the sensing target needs to transmit dedicated probing signals to the receiver actively.
With the growing trend of integrating sensing and communication functionalities into a shared antenna array in 6G \cite{liu2022integrated}, passive sensing is also worth investigating, which alternatively utilizes the reflected probing signal to position the targets themselves.
Hence, passive sensing can exempt the need for dedicated probing signals from the target.
To enable CAPA-based passive NISE systems, three key challenges must be addressed:
\textbf{First, a positioning algorithm is required for CAPA-based passive NISE scenarios.}
Compared to active sensing, passive sensing lacks dedicated probing signals from the target, making accurate positioning more challenging.
\textbf{Second, a sensing performance bound specifically for CAPA-based NISE scenarios needs to be derived, based on which the sensing performance can be quantified.}
Moreover, this sensing performance bound needs to be derived based on a continuous source function and must be tractable to allow for further optimization.
\textbf{Third, a sensing-performance-bound optimization algorithm needs to be devised to enhance the sensing accuracy.}
It is noteworthy that, in contrast to discrete-variable-based optimizations in conventional SPDAs, CAPA-based passive sensing requires optimizing a continuous source current.
Currently, the most widely adopted method is the wavenumber-domain method \cite{zhang2023pattern}, which discretizes the continuous source current function by sampling a finite number of points in the wavenumber domain.
By doing so, instead of directly dealing with a continuous function, Fourier coefficients are optimized equivalently, whose number is proportional to the carrier frequency.
Although this approach is straightforward, its complexity is prohibitive due to the high volume of Fourier coefficients, particularly for planar arrays operating at high frequencies.
Therefore, dealing with the three challenges of affordable complexity plays a fundamental role in CAPA-based passive NISE systems.

To bridge the knowledge gap, this paper aims to address the three key challenges by deriving a performance bound for multi-target NISE scenarios and proposing a manifold-based optimization method to enhance sensing accuracy.
Our main contributions are summarized as follows:
\begin{itemize}
    \item We model the round-trip channel for CAPA-based NISE based on EM theories, where the base station (BS) is equipped with two separate Tx and Rx CAPAs to position the multiple targets in a passive sensing fashion. 

    \item We propose an MLE-based method to estimate the positions of multiple targets based on EM-based round-trip channel modeling.
    This method is developed under the maximum likelihood rule.
    
    \item We derive the CRB for the above setup, providing a mathematically tractable lower bound for sensing performance. 
    Based on the derived CRB, a CRB minimization problem is formulated under a unit power constraint.
    By solving this problem, the sensing accuracy can be improved.

    \item We prove that the optimal solution resides in a subspace spanned by the array responses of all the sensing targets.
    Subsequently, we propose a low-complexity subspace manifold gradient descent (SMGD) to solve the CRB minimization problem.

    \item We present numerical results to verify the performance of the proposed SMGD method and provide insights into the CRB in CAPA-based systems.
    The simulation results demonstrate that: 
    i)~The proposed SMGD algorithm can solve the CRB minimization problem effectively at a reduced cost of computational complexity. 
    ii)~The achieved CRB of CAPAs outperforms that of its SPDA counterpart, confirming CAPA's advanced sensing capabilities.
    
\end{itemize}
\subsection{Organization and Notations}
The organizations of this paper are showcased as follows.
Section \ref{sect:system model} presents the system model and the passive multi-target sensing signal model.
The CRB and MLE spectra derivations are detailed in Section \ref{sect:prob_form}.
Based on the derivations in the former section, the CRB minimization problem is formulated and solved by the proposed SMGD algorithm in Section \ref{sect:crb_opti}.
Numerical results are elaborated in Section \ref{section: simulation} to demonstrate the performance of the proposed algorithm.
Finally, the conclusion is drawn in Section \ref{sect:conclu}.

\textit{Notations:} 
Scalars are denoted using regular typeface, while vectors and matrices are denoted by bold-face letters.
$\Re\{(\cdot)\}$ and $\Im\{(\cdot)\}$ are the operations to extract the real and imaginary parts of $(\cdot)$.
$\mathrm{Tr}\{(\cdot)\}$, $(\cdot)^\mathrm{T}$, $(\cdot)^*$, $[(\cdot)]^{-1}$, and $\mathrm{Rank}\{\cdot\}$ represent the trace, transpose, conjugate, matrix inverse, and calculating-rank operations, respectively.
$\mathbf{B}^{\mathrm{Tr}}$ and $\otimes$ denote the blockwise trace operation for a block matrix $\mathbf{B}$ and the Kronecker product, respectively.
$\langle\cdot, \cdot\rangle$ denotes the inner product between two continuous functions.
$\mathbb{R}^{M \times N}$ and $\mathbb{C}^{M \times N}$ denote the $M \times N$ real and imaginary matrix spaces, respectively.
$[(\cdot)]_{m,n}$ means to take the $(m,n)$-th entry of matrix/vector $(\cdot)$. 
${\mathrm{j}}= \sqrt{-1}$ is the imaginary unit.

\section{System Model} \label{sect:system model} 
\begin{figure}[t!]
    \centering
    \includegraphics[width=0.7\linewidth]{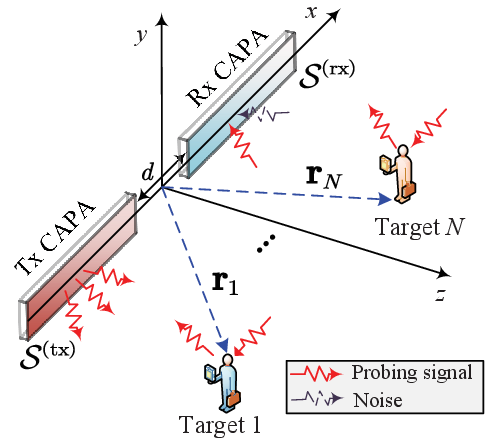}
    \caption{An illustration of the CAPA-based NISE system setup.}
    \label{fig:syst_model}
\end{figure}
The system model of this work is presented in Fig. \ref{fig:syst_model}, where the BS is equipped with a Tx planar CAPA (P-CAPA) for transmission and an Rx P-CAPA for reception.
There are $N$ sensing targets being positioned by the BS.
Without loss of generality, we assume that the Rx and Tx P-CAPAs are placed on the $XOY$ plane, separated by a distance of $d$. 
The origin of the coordinate system is set at the center point between Rx and Tx P-CAPAs.
It is noted that, due to the adoption of P-CAPAs, channel responses across the $z$-axis are negligible \cite{zhao2025continuous}. 
Further, we assume the targets are located within the radiating near-field region of the BS, which is defined by $r \le \frac{2D^2}{\lambda}$ with $D$ and $\lambda$ denoting the aperture size and wavelength, respectively.
For the BS, the sizes of the Tx and Rx P-CAPAs are denoted by $H_{}^{\left( \mathrm{tx} \right)} \times W_{}^{\left( \mathrm{tx} \right)}$ and $H_{}^{\left( \mathrm{rx} \right)} \times W_{}^{\left( \mathrm{rx} \right)}$ respectively, where $H_{}^{\left( \mathrm{\cdot} \right)}$ and $W_{}^{\left( \mathrm{\cdot} \right)}$ denote the height and the width of the Rx and Tx P-CAPAs, respectively.
The coordinate regions of the Tx and Rx P-CAPAs are specified by
\begin{align}
    \mathcal{S} ^{\left( \mathrm{tx} \right)}&=\left\{ \mathbf{p} \triangleq [p^{\left( x \right)}, p^{\left( y \right)}, p^{\left( z \right)} ]^{\mathrm{T}}:W_{\min}^{\left( \mathrm{tx} \right)}\le p^{\left( x \right)}\le W_{\max}^{\left( \mathrm{tx} \right)}, \right. \notag \\ &\qquad \qquad \qquad \left. H_{\min}^{\left( \mathrm{tx} \right)}\le p^{\left( y \right)}\le H_{\max}^{\left( \mathrm{tx} \right)},~~p^{\left( z \right)}=0 \right\}, \\
    \mathcal{S} ^{\left( \mathrm{rx} \right)}&=\left\{ \mathbf{q}\triangleq [q^{\left( x \right)}, q^{\left( y \right)}, q^{\left( z \right)} ]^{\mathrm{T}}:W_{\min}^{\left( \mathrm{rx} \right)}\le q^{\left( x \right)}\le W_{\max}^{\left( \mathrm{rx} \right)}, \right. \notag \\ &\qquad \qquad \qquad  \left. 
    H_{\min}^{\left( \mathrm{rx} \right)}\le q^{\left( y \right)}\le H_{\max}^{\left( \mathrm{rx} \right)},~~q^{\left( z \right)}=0 \right\},
\end{align}
where the boundaries are calculated by $W_{\min}^{\left( \mathrm{tx} \right)}=-W_{}^{\left( \mathrm{tx} \right)}-d/2$, 
$W_{\max}^{\left( \mathrm{tx} \right)}=-d/2$, $W_{\min}^{\left( \mathrm{rx} \right)}=d/2$, $W_{\max}^{\left( \mathrm{rx} \right)}=W_{}^{\left( \mathrm{rx} \right)}+d/2$, $H_{\min}^{\left( \mathrm{tx} \right)}= -0.5 H^{\mathrm{(tx)}}$, $H_{\max}^{\left( \mathrm{tx} \right)}= +0.5 H^{\mathrm{(tx)}}$, 
$H_{\min}^{\left( \mathrm{rx} \right)}= -0.5 H^{\mathrm{(rx)}}$, and $H_{\max}^{\left( \mathrm{rx} \right)}= +0.5 H^{\mathrm{(rx)}}$, respectively.
The position of the $n$-th ST is specified by $\mathbf{r}_n \triangleq [r^{\left( x \right)}_n, r^{\left( y \right)}_n, r^{\left( z \right)}_n]^{\mathrm{T}}\in \mathbb{R}^{3 \times 1}$, where $n \in \mathcal{N}$ and $ \mathcal{N} \triangleq \{1,2,..., N\}$.

\subsection{Derivation of Channel Response}
Let $\mathbf{p} \in \mathbb{R}^{3 \times 1}$ be a general point in $\mathbb{R}^3$ space.
Considering the source current distribution $\mathbf{J}(\mathbf{p},w) \in \mathbb{R}^{3 \times 1}$ with a frequency of $w$, the resultant scalar electric field $\mathbf{E}(\mathbf{p},w) \in \mathbb{R}^{3 \times 1}$ excited by this source, in which $w$ denotes the angular frequency.
According to Maxwell equations \cite{jackson1998classical}, the following derivations can be performed:
\begin{align}
    &\nabla \times \left( \nabla \times \mathbf{E}\left( \mathbf{p},w \right) \right) =-\mathrm{j}w\mu \nabla \times \mathbf{H}\left( \mathbf{p},w \right) 
    \notag \\
    &=-\mathrm{j}w\mu \left( \mathbf{J}\left( \mathbf{p},w \right) +\mathrm{j}w\epsilon \mathbf{E}\left( \mathbf{p},w \right) \right) 
    \notag \\
    &=-\mathrm{j}w\mu \mathbf{J}\left( \mathbf{p},w \right) +w^2\mu \epsilon \mathbf{E}\left( \mathbf{p},w \right) 
    \notag \\
    &=-\mathrm{j}k_0\eta _0\mathbf{J}\left( \mathbf{p},w \right) +k_{0}^{2}\mathbf{E}\left( \mathbf{p},w \right) 
    \notag \\
    &\Rightarrow \left( -\nabla \times \nabla \times +k_{0}^{2} \right) \mathbf{E}\left( \mathbf{p},w \right) =\mathrm{j}k_0\eta _0\mathbf{J}\left( \mathbf{p},w \right), \label{eq:equation_maxwell}
\end{align}
where $k_0\triangleq w\sqrt{\mu \epsilon}$ and $\eta _0 \triangleq \sqrt{\mu /\epsilon}$ denote the wavenumber and wave impedance in free space with $\mu$ and $\epsilon$ being the permeability and the permittivity of the medium, respectively.
Under the narrowband assumption \cite{r3_8}, the frequency term $w$ is therefore fixed, making the whole system spatially linear and spatial-invariant.
In addition, the solution to \eqref{eq:equation_maxwell} can be expressed as 
\begin{align}
    \mathbf{E}\left( \mathbf{\mathbf{r}} \right) =\int_{\mathbb{R} ^3}{\mathbf{G}\left( \mathbf{r},\mathbf{p} \right) \mathbf{J}\left( \mathbf{p} \right) \mathrm{d}\mathbf{p}}, 
\end{align}
where kernel $\mathbf{G}\left( \mathbf{k},\mathbf{p} \right) \in \mathbb{C}^{3 \times 3}$ is the vector Green's function.
The vector Green's function can be expressed as
\begin{align}
    &\mathbf{G}\left( \mathbf{r},\mathbf{p} \right) =\frac{\mathrm{j}k_0\eta _0e^{-\mathrm{j}k_0\left\| \mathbf{k} \right\|}}{4\pi \left\| \mathbf{k} \right\|}\left( \left( \mathbf{I}_3-\hat{\mathbf{k}}\hat{\mathbf{k}}^{\mathrm{H}} \right) \right. \notag \\
    &\quad \quad \quad  \left.+\frac{\mathrm{j}k_{0}^{-1}}{\left\| \mathbf{k} \right\|}\left( \mathbf{I}_3-3\hat{\mathbf{k}}\hat{\mathbf{k}}^{\mathrm{H}} \right) -\frac{k_{0}^{-2}}{\left\| \mathbf{p} \right\| ^2}\left( \mathbf{I}_3-3\hat{\mathbf{k}}\hat{\mathbf{k}}^{\mathrm{H}} \right) \right), \label{eq:origin_green_func}
\end{align}
where $\mathbf{k}\triangleq \mathbf{r}-\mathbf{p}$ and $\hat{\mathbf{k}}\triangleq \mathbf{k}/\left\| \mathbf{k} \right\|$.
According to \cite{wan2024near}, the vector Green’s function in \eqref{eq:origin_green_func} consists of three terms: the first represents the radiating near- and far-field components, while the second and third correspond to the reactive near-field region. As noted in \cite{ouyang2024on}, the reactive near-field region extends only a few wavelengths and can thus be neglected. Numerically, under a typical near-field setup in \cite{li2024near}, the power ratios of the second and third terms are approximately $10^{-5}$ and $10^{-9}$, respectively.
Therefore, the vector Green's function can be simplified as
\begin{align}
    \mathbf{G}\left( \mathbf{r},\mathbf{p} \right) \approx \frac{\mathrm{j}k_0\eta _0e^{-\mathrm{j}k_0\left\| \mathbf{k} \right\|}}{4\pi \left\| \mathbf{k} \right\|}\left( \mathbf{I}_3-\hat{\mathbf{k}}\hat{\mathbf{k}}^{\mathrm{H}} \right). \label{eq:vectorized_round_trip_channel}
\end{align}
Furthermore, according to \cite{wan2024near}, we have $\mathrm{Tr}\{( \mathbf{I}_3 -  \hat{\mathbf{k}}\hat{\mathbf{k}}^{\mathrm{H}})( \mathbf{I}_3 -  \hat{\mathbf{k}}\hat{\mathbf{k}}^{\mathrm{H}})\} = \mathrm{Tr}\{\mathbf{I}_3 - \hat{\mathbf{k}}\hat{\mathbf{k}}^{\mathrm{H}}\} = 3 - \hat{\mathbf{k}}^{\mathrm{H}}\hat{\mathbf{k}}^{}=2$, indicating that the average power of EM field is not relevant to the direction of $\hat{\mathbf{k}}$ if the power of $\mathbf{J}(\mathbf {p})$ is equally distributed in all polarization directions. 
As such, the vector Green's function can be converted to the following scalar form:
\begin{align}
    g\left( \mathbf{r},\mathbf{p} \right) =\frac{\mathrm{j}k_0\eta _0e^{-\mathrm{j}k_0\left\| \mathbf{k} \right\|}}{4\pi \left\| \mathbf{k} \right\|}.
\end{align}
Corresponding to the following scalar form, the excited electric field and the source current can be written as scalars, i.e., $E(\mathbf{r})$ and $J(\mathbf{p})$.
In particular, this relationship is given by 
\begin{align}
    E\left( \mathbf{\mathbf{r}} \right) =\int_{\mathcal{S} \in \mathbb{R} ^2}{g\left( \mathbf{r},\mathbf{p} \right) J\left( \mathbf{p} \right) \mathrm{d}\mathbf{p}}, 
\end{align}
where $\mathcal{S}$ denotes the transmit plane. 
Here, the single-trip channel response of CAPA is derived from EM theories.
However, in sensing setups, we need to consider both the probing direction, i.e., Tx-CAPA to sensing targets (STs), and the reflection direction, i.e., STs to Rx-CAPA, which will complicate the analysis based on EM theories.
However, thanks to the introduction of Green's function, our derivations for the round-trip channel can be conducted concerning the operator, i.e., Green's function.
    
\subsection{Green's Function-Based Round Channel Modeling}
\begin{figure}[t!]
    \centering
    \includegraphics[width=0.45\linewidth]{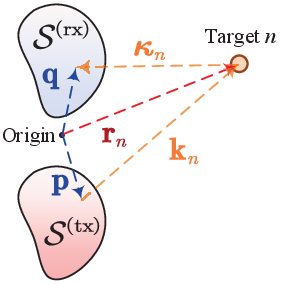}
    \caption{An explanation of geometry for the $n$-th ST.}
    \label{fig:vect_rela}
\end{figure}

Based on the channel response derived in the former sub-section, we will present the modeling of the round-trip channel, i.e., Tx-CAPA $\rightarrow$ STs $\rightarrow$ Rx-CAPA. 
In particular, the round-trip channel consists of both the probing and the echo spatial propagation model and the reflecting model between them.
As illustrated in Fig. \ref{fig:vect_rela}, for the $n$-th ST, we define the transmit vector as $\mathbf{k}_n\triangleq \mathbf{r}_n -\mathbf{p}$ and the receive vector $\boldsymbol{\kappa}_n \triangleq \mathbf{q} - \mathbf{r}_n $, given that the coordinates $\mathbf{p}\in \mathcal{S}^{\mathrm{(tx)}}$ and $\mathbf{q}\in \mathcal{S}^{\mathrm{(rx)}}$ are of our interest.
Thus, the probing and receiving channel response modeled by Green's function can be expressed as scalar Green's functions, i.e., 
\begin{align}
    a_{\mathrm{t}}\left( \mathbf{n},\mathbf{p} \right) &\triangleq g\left( \mathbf{n},\mathbf{p} \right) =\frac{\mathrm{j}\eta _0k_0\mathrm{e}^{-\mathrm{j}k_0\left\| \mathbf{n}-\mathbf{p} \right\|}}{{4\pi}\left\| \mathbf{n}-\mathbf{p} \right\|}, \label{eq:transmit_response} \\
    a_{\mathrm{r}}\left( \mathbf{q},\mathbf{m} \right) &\triangleq g\left( \mathbf{q},\mathbf{m} \right) =-\frac{\mathrm{j}\eta _0k_0\mathrm{e}^{\mathrm{j}k_0\left\| \mathbf{q}-\mathbf{m} \right\|}}{{4\pi}\left\| \mathbf{q}-\mathbf{m} \right\|}. \label{eq:receive_response}
    \end{align}
Then, in light of \cite{poon2005degrees}, the coupling effects between responses $a_{\mathrm{t}}\left( \mathbf{n},\mathbf{p} \right)$ and $a_{\mathrm{r}}\left( \mathbf{q},\mathbf{m} \right)$ can be modeled by the product of two Dirac functions, i.e.,
\begin{align}
       c(\mathbf{n},\mathbf{r},\mathbf{m})=\frac{1}{\sqrt{N}}\sum\nolimits_{n=1}^N{\alpha _n\delta (\mathbf{r}_n-\mathbf{n})\delta \left( \mathbf{r}_n-\mathbf{m} \right)}, \label{eq:scatter}
\end{align}
where ${\alpha }_{n\,\,} \in \mathbb{C}$ denotes the reflection coefficient of the $n$-th ST \cite{liu2022integrated}, and $\delta(\cdot)$ is the Dirac function.
Then, jointly considering \eqref{eq:transmit_response}, \eqref{eq:receive_response}, and \eqref{eq:scatter}, the round-trip channel can be derived by 
\begin{align}
     h\left( \mathbf{q},\mathbf{p} \right) &=\frac{1}{\sqrt{N}}{\int_{\mathbb{R}^{3}}{g\left( \mathbf{q},\mathbf{n} \right) c(\mathbf{n},\mathbf{r},\mathbf{m})g\left( \mathbf{m},\mathbf{p} \right) \mathrm{d}\mathbf{n}\mathrm{d}}\mathbf{m}}
    \notag\\
    &=\frac{1}{\sqrt{N}}{\int_{\mathbb{R}^{3}}{\sum\nolimits_{n=1}^N{{\frac{\mathrm{j}\eta _0k_0e^{-\mathrm{j}k_0\left\| \mathbf{n}-\mathbf{p} \right\|}}{\sqrt{4\pi}\left\| \mathbf{n}-\mathbf{p} \right\|}\alpha _n}}}} \notag \\
    &\quad \times {{{ \delta (\mathbf{r}_n-\mathbf{n}) \delta \left( \mathbf{r}_n-\mathbf{m} \right) \frac{\mathrm{j}\eta _0k_0e^{-\mathrm{j}k_0\left\| \mathbf{q}-\mathbf{m} \right\|}}{\sqrt{4\pi}\left\| \mathbf{q}-\mathbf{m} \right\|}\mathrm{d}\mathbf{n}\mathrm{d}\mathbf{m}}}} \notag
    \\
    &\overset{(\mathrm{a)}}{=}\frac{1}{\sqrt{N}}c_0\sum\nolimits_{n=1}^N{\alpha _n\frac{e^{-\mathrm{j}k_0\left\| \mathbf{q}-\mathbf{r}_n \right\|}}{\left\| \mathbf{q}-\mathbf{r}_n \right\|}\frac{e^{-\mathrm{j}k_0\left\| \mathbf{r}_n-\mathbf{p} \right\|}}{\left\| \mathbf{r}_n-\mathbf{p} \right\|}}, \label{eq:round-trip_channel}
\end{align}
where $c_0\triangleq \frac{\eta _{0}^{2}k_{0}^{2}}{16\pi^2 \sqrt{N\,\,}}$ and $\mathrm{(a)}$ is obtained using the property of Dirac function.

In the sequel, we will present the noise model.
Particularly, we assume that the noise term $n(\mathbf{q}) \in \mathbb{C}$ with $\mathbf{q} \in \mathcal{S}^{(\mathrm{rx})}$ follows a spatially uncorrelated zero-mean complex Gaussian process as described in \cite{amico2022cramer} and \cite{jensen2008capacity}.
The correlation function of such a process is given by 
\begin{align}
    \mathbb{E}\{n(\mathbf{q}) n^{*}(\mathbf{q}^\prime)\} = {\sigma_0^2} \delta(\mathbf{q}-\mathbf{q}^\prime), \label{eq:noise_correlation}
\end{align}
where $\sigma_0^2$ denotes the power angular density measured by $\mathrm{V}^2$, where $\mathrm{V}$ represents volts.
Here, a spatially white noise model is adopted for simplicity. For those interested in spatially colored noise modeling, further details can be found in \cite{qian2024on}.
It is worth noting that the term $\delta(\mathbf{q}-\mathbf{q}^\prime)$ in \eqref{eq:noise_correlation} reveals that the noise is uncorrelated across different points $\mathbf{q}$ and $\mathbf{q}^\prime$ on Rx P-CAPA.

On top of the above, due to the presence of noise, the received echo signal $y(\mathbf{q})$ at Rx P-CAPA can be expressed as
\begin{align}
    y\left( \mathbf{q} \right) ={E}(\mathbf{q})+n\left( \mathbf{q} \right), \label{eq:conti}
\end{align}
where the excited electric filed ${E}(\mathbf{q})$ is defined by
\begin{align}
     {E}(\mathbf{q}) \triangleq \int_{\mathcal{S} ^{(\mathrm{tx)}}}{h\left( \mathbf{q},\mathbf{p} \right) {J}(\mathbf{p}) \mathrm d}\mathbf{p}. \label{eq:conti_ele_field}
\end{align}
It is noted that the excited electric field in \eqref{eq:conti_ele_field} is the response at an infinitesimal point on CAPA incurred by the joint effect of the source current on the whole transmit plane.
In discrete MIMO systems, the response of an arbitrary antenna is a summation of several discrete links connecting the source, while in the context of CAPA, this summation is converted into an integral. 
Moreover, when the whole receiving plane of CAPA is analyzed, we need to perform integration of ${E}(\mathbf{q}) $ of every point.
As mentioned in the former section, we assume that the power of the source current is evenly distributed to the three orthogonal polarization directions.
As such, the vector source current function can be secularized. 
When a particular polarization direction is considered, say $\tilde{\mathbf{v}}$ at the the transmit CAPA and $\tilde{\mathbf{v}}$ with $\|\tilde{\mathbf{v}}\|=1$ and $\|\tilde{\mathbf{u}}\|=1$, the projection onto the propagation direction is needed.
In this case, the scalar current and the excited electric field can be respectively given by $J(\mathbf{p}) \triangleq \left<  \tilde{\mathbf{v}}, \mathbf J(\mathbf{p})\right>$ and $E(\mathbf{q}) \triangleq \left<  \tilde{\mathbf{u}}, \mathbf E(\mathbf{q})\right>$.

In the considered CAPA-based passive NISE scenario, the Tx P-CAPA transmits probing signals with a designed source current function ${J}(\mathbf{p})$ over region $\mathcal{S}^{(\mathrm{tx})}$.
After the probing signal is reflected by the STs, the Rx P-CAPA receives the echo signal ${E}(\mathbf{q})$ received over region $\mathcal{S}^{(\mathrm{rx})}$.
Therefore, we aim to extract the targets' positional information $\{\mathbf{r}_i\}_{i=1,...,N}$ from the received echo signal ${E}(\mathbf{q})$ over the entire Tx P-CAPA $\mathcal{S}^{\mathrm{(rx)}}$.
Since $\{\mathbf{r}_i\}_{i=1,...,N}$ is of interest, we mainly focus on the positioning accuracy.
\begin{remark} \emph{(\emph{Difference Between CAPA and SDPA in MIMO scenario}): Due to the discrete placement of antenna elements at transceivers, MIMO channels in SDPA are two-dimensional matrices, with each entry representing a specific propagation method. 
In addition, when signals are conveyed through a wireless channel, matrix-based channel modeling induces summations to reflect the input-output relationships.
However, in CAPA, the antenna aperture contains an infinite number of antennas, which cannot be rewritten as discrete entities, thus leading to infinite dimensions of the wireless channel.
Therefore, the MIMO channel in CAPA can only be written as a scalar equation, describing the excited electric field (infinitesimal) and its source current (infinitesimal).
Moreover, the input-output relationship in CAPA is depicted by integrals rather than summations, as a result of the continuous placement of antennas.}
\end{remark}

\section{Derivations of MLE and CRB} \label{sect:prob_form}
In this section, we first present the derivation of the MLE method, which is a positioning method for multiple targets under the maximum likelihood rule.
Then, to characterize the sensing performance, we derive CRB as the performance metric, which is widely used in literature \cite{liu2022cramer, song2023intelligent, liu2022integrated}.

\subsection{MLE Derivations for CAPAs} \label{sect:mle}
Given that the positions and the reflection coefficients of the $N$ STs need to be estimated, the unknown parameter vector $ \boldsymbol{\xi }$ can be expressed as 
\begin{align}
    \boldsymbol{\xi }&=\left[ \mathbf{r}_1,...,\mathbf{r}_N,\Re \left\{ \alpha _1 \right\} ,\Im \left\{ \alpha _1 \right\} ,...,\Re \left\{ \alpha _N \right\} ,\Im \left\{ \alpha _N \right\} \right] ^{\mathrm{T}}\notag \\
    &=\left[ \mathbf{r},\boldsymbol{\alpha } \right] ^{\mathrm{T}}\in \mathbb{R} ^{5N\times 1},
\end{align}
where $\mathbf{r}\triangleq \left[ \mathbf{r}_{1},...,\mathbf{r}_{N} \right]^{\mathrm{T}} 
$ contains the positions of the $N$ STs and $\boldsymbol{\alpha }\triangleq \left[ \Re \left\{ \alpha _1 \right\} ,\Im \left\{ \alpha _1 \right\} ,...,\Re \left\{ \alpha _N \right\} ,\Im \left\{ \alpha _N \right\} \right] ^{\mathrm{T}}$ contains the real and imaginary parts of the reflection coefficients of the $N$ STs.
For target positioning, $\mathbf{r}$ can be estimated using classic MLE based on received observation $y(\mathbf{q})$ for $\mathbf{q} \in \mathcal{S}^{\mathrm{(rx)}}$.
It is noted that, compared to conventional SPDA-based sensing, the derivation of MLE in this work is based on a continuous source current function.
Based on the received echo signals denoted by $y(\mathbf{q})$, the estimated version of the unknown position vector $\hat{\mathbf{r}}$ can be obtained by maximizing the likelihood.
The discrete expression log-likelihood function can be found in \cite{amico2022cramer}. 
To derive the likelihood function, we first consider a discrete case. 
In particular, we discretize the Rx P-CAPA as grids indexed by $i_x\in\mathcal{I}_x$ and $i_y\in\mathcal{I}_y$.
    Here, the set cardinalities $|\mathcal{I}_x|$ and $|\mathcal{I}_y|$ denote the number of samples along $x$-axis and $y$-axis, respectively.
    Therefore, regarding the real received signal $y\left( \mathbf{q} \right)$ and estimated electric field  $\tilde{E}_n\left( \mathbf{q} \right)$ for $n\in\mathcal{N}$ on an arbitrary grid denoted by $\mathcal{S}_{i_x, i_y}^{\mathrm{(rx)}}$, we have the following derivations:
    \begin{align}
        &\frac{1}{\sqrt{\Delta A}}\int_{\mathcal{S} _{i_x,i_y}^{\left( \mathrm{rx} \right)}}{y\left( \mathbf{q} \right) \mathrm{d}\mathbf{q}=\frac{1}{\sqrt{\Delta A}}y\left( \mathbf{q}_{i_x,i_y} \right) \Delta A}\notag \\
        &=y\left( \mathbf{q}_{i_x,i_y} \right) \sqrt{\Delta A}\notag \\
        &\frac{1}{\sqrt{\Delta A}}\int_{\mathcal{S} _{i_x,i_y}^{\left( \mathrm{rx} \right)}}{\sum_{n=1}^N{\alpha _n\tilde{E}_n\left( \mathbf{q} \right) \mathrm{d}\mathbf{q}}}=\sum_{n=1}^N{\alpha _n\tilde{E}_n\left( \mathbf{q}_{i_x,i_y} \right) \sqrt{\Delta A}}, \notag
    \end{align}
    where $\mathbf{q}_{i_x,i_y}$ is the position of an arbitrary grid cell on Rx CAPA, and $\Delta A$ denotes the area of each grid cell.
    Building on this, the discrete likelihood function is formulated as
    \begin{align}
        \mathcal{L} _{\mathrm{d}}&\triangleq \sum_{i_x\in \mathcal{I} _x,i_y\in \mathcal{I} _y}{\left| y\left( \mathbf{q}_{i_x,i_y} \right) \sqrt{\Delta A}-\sum_{n=1}^N{{\alpha}_n\tilde{E}_n\left( \mathbf{q}_{i_x,i_y} \right) \sqrt{\Delta A}} \right|^2} \notag \\
        &=\sum_{i_x\in \mathcal{I} _x,i_y\in \mathcal{I} _y}{\left| y\left( \mathbf{q}_{i_x,i_y} \right) -\sum_{n=1}^N{{\alpha}_n\tilde{E}_n\left( \mathbf{q}_{i_x,i_y} \right)} \right|^2 \Delta A}, \notag
    \end{align}
    where $ \Delta A > 0$.
    Then, to derive the continuous form of $\mathcal{L}_{\rm d}$, we perform the following derivations:
    \begin{align}
        \mathcal{L} &\triangleq \lim_{\Delta A \rightarrow 0} \mathcal{L} _{\mathrm{d}} \notag \\
        &=\int_{H_{\min}^{\left( \mathrm{rx} \right)}}^{H_{\max}^{\left( \mathrm{rx} \right)}}{\int_{W_{\min}^{\left( \mathrm{rx} \right)}}^{W_{\max}^{\left( \mathrm{rx} \right)}}{\left| y\left( \mathbf{q} \right) -\sum_{n=1}^N{{\alpha}_n\tilde{E}_n\left( \mathbf{q} \right)} \right|^2\mathrm{d}x\mathrm{d}y}}.  \notag 
    \end{align}
    Defining $\mathbf{{E}}(\mathbf{q})\triangleq \left[ \tilde{E}_1(\mathbf{q}),...,\tilde{E}_N(\mathbf{q}) \right] ^{\mathrm{T}}$ and $\boldsymbol{\alpha }\triangleq \left[ \alpha _1,...,\alpha _N \right] ^{\mathrm{T}}$, $\mathcal{L}$ is compactly given by
    \begin{align}
    \mathcal{L} 
    &= \int_{H_{\min}^{(\mathrm{rx})}}^{H_{\max}^{(\mathrm{rx})}} 
        \int_{W_{\min}^{(\mathrm{rx})}}^{W_{\max}^{(\mathrm{rx})}} 
        \left| y\left( \mathbf{q} \right) - \mathbf{E}^{\mathrm{T}}\left( \mathbf{q} \right) \boldsymbol{\alpha} \right|^2 \, \mathrm{d}x \, \mathrm{d}y \notag \\
    &= \int_{H_{\min}^{(\mathrm{rx})}}^{H_{\max}^{(\mathrm{rx})}}
   \int_{W_{\min}^{(\mathrm{rx})}}^{W_{\max}^{(\mathrm{rx})}}
   \left( |y(\mathbf q)|^2
          - 2\,\Re\!\left\{ y^*(\mathbf q)\,\mathbf E^{\mathrm T}(\mathbf q)\,\boldsymbol\alpha \right\}
   \right. \notag \\
    &\qquad\left.
       + \boldsymbol\alpha^{\mathrm H}\,\mathbf E^*(\mathbf q)\,\mathbf E^{\mathrm T}(\mathbf q)\,\boldsymbol\alpha
       \right)\, \mathrm dx\,\mathrm dy .
    \end{align}
    To find the optimal solution for estimating $\alpha_n$ for $n\in\mathcal{N}$, we calculate the partial derivative with respect to $\boldsymbol{\alpha}^*$ and let this derivative equals zero, thus yielding
    \begin{align}
       &\frac{\partial \mathcal{L}}{\partial \boldsymbol{\alpha}^*}=-\mathbf{m} +\mathbf{P}\boldsymbol{\alpha }=0,
        \label{A-2}
    \end{align}
    where 
    \begin{align}
        &I_1\triangleq \int_{\mathcal{S} ^{(\mathrm{rx)}}}{\left| y\left( \mathbf{q} \right) \right|^2\mathrm{d}\mathbf{q}}, \notag \\
        &\mathbf{m}^{\mathrm{H}}\triangleq \int_{\mathcal{S} ^{(\mathrm{rx)}}}{y^*\left( \mathbf{q} \right) \mathbf{E}_{}^{\mathrm{T}}\left( \mathbf{q} \right) \mathrm{d}\mathbf{q}}=I_2, \\
        &\mathbf{P}\triangleq \int_{\mathcal{S} ^{(\mathrm{rx)}}}{\mathbf{E}_{}^{*}\left( \mathbf{q} \right) \mathbf{E}_{}^{\mathrm{T}}\left( \mathbf{q} \right) \mathrm{d}\mathbf{q}} 
    \end{align}
    Here, mathematical derivations lead us to the following linear system:
    \begin{align}
        \mathbf{P}\boldsymbol{\alpha }=\mathbf{m}, \label{A-4}
    \end{align}
    whose solution can be expressed as
    \begin{align}
        \boldsymbol{\alpha }^\star=\mathbf{P}^{-1}\mathbf{m}.\label{A-6}
    \end{align}
    Therefore, by substituting the reflection coefficients with the results in \eqref{A-6}, the likelihood function can be written as
   \begin{align}
        \mathcal{L} ^{\prime}&=I_1-\mathbf{m}^{\mathrm{H}}\boldsymbol{\alpha }^{}-\boldsymbol{\alpha }^{\mathrm{H}}\mathbf{m}^{}+\boldsymbol{\alpha }^{\mathrm{H}}\mathbf{P}\boldsymbol{\alpha }^{}\notag
        \\
        &=I_1-\mathbf{m}^{\mathrm{H}}\mathbf{P}^{-1}\mathbf{m}-\mathbf{m}^{\mathrm{H}}\left( \mathbf{P}^{-1} \right) ^{\mathrm{H}}\mathbf{m}\notag \\
        &\qquad \qquad \quad +\mathbf{m}^{\mathrm{H}}\left( \mathbf{P}^{-1} \right) ^{\mathrm{H}}\mathbf{PP}^{-1}\mathbf{m}\notag 
        \\
        &=I_1-\mathbf{m}^{\mathrm{H}}\mathbf{P}^{-1}\mathbf{m}.
    \end{align}
    Furthermore, by excluding constant value produced by $\mathrm{I}_1$ and adding a minus sign, we have 
    \begin{align}
        \mathcal{L} ^{\prime}=\mathbf{m}^{\mathrm{H}}\mathbf{P}^{-1}\mathbf{m}. \label{A-9}
    \end{align}
    Therefore, by maximizing \eqref{A-9}, the position of the target can be identified.
    Specifically, for the single target scenario, i.e., $N=1$, we have $m=\int_{\mathcal{S} ^{(\mathrm{rx)}}}{\tilde{E}^*\left( \mathbf{q} \right) y^{}\left( \mathbf{q} \right) \mathrm{d}\mathbf{q}}$ and $\rho =\int_{\mathcal{S} ^{(\mathrm{rx)}}}{\left| \tilde{E}\left( \mathbf{q} \right) \right|^2\mathrm{d}\mathbf{q}}$.
    Then, the optimal reflection coefficient can be computed as $\alpha^\star = m/\rho$, which further yields 
    \begin{align}
         \mathcal{J} ^{\prime}= \frac{\left|m\right|^2}{\rho} = \frac{\left| \int_{\mathcal{S} ^{(\mathrm{rx)}}}{\tilde{E}_{}^{*}(\mathbf{q})y(\mathbf{q})\mathrm{d}\mathbf{q}} \right|^2}{\int_{\mathcal{S} ^{(\mathrm{rx)}}}{\left| \tilde{E}(\mathbf{q}) \right|^2\mathrm{d}\mathbf{q}}}.
    \end{align}
Hence, the MLE spectrum can be obtained by traversing all possible $\mathbf{r}$ \cite{song2023intelligent}.
The positions of the STs can be found at the spike locations on the MLE spectrum.

\subsection{CRB Derivation for CAPAs}
In this subsection, we characterize the sensing performance lower bound using CRB.
To derive CRB, we first calculate the Fisher information matrix (FIM) for estimating $\boldsymbol{\xi }$, which is given by a partitioned matrix as:
\begin{align}
    \mathbf{F}_{\boldsymbol{\xi }}=\left[ \begin{matrix}
	\mathbf{F}_{\mathbf{rr}}&		\mathbf{F}_{\mathbf{r}\tilde{\boldsymbol{\alpha}}}\\
	\mathbf{F}_{\mathbf{r}\tilde{\boldsymbol{\alpha}}}^{\mathrm{T}}&		\mathbf{F}_{\tilde{\boldsymbol{\alpha}}\tilde{\boldsymbol{\alpha}}}\\
\end{matrix} \right]. \label{eq:j-cont}
\end{align}
According to \cite{amico2022cramer, chen2024cramer}, each block matrix in FIM $\mathbf{F}_{\boldsymbol{\xi }}$ can be calculated according to 
\begin{align}
    \left[ \mathbf{F}_{\mathbf{rr}} \right] _{m,n}&=\frac{2}{\sigma ^2}\int_{\mathcal{S} ^{(\mathrm{rx)}}}{\Re \left\{ \frac{\partial {E}\left( \mathbf{q} \right)}{\partial \left[ \mathbf{r} \right] _m}\frac{\partial {E}^*\left( \mathbf{q} \right)}{\partial \left[ \mathbf{r} \right] _n} \right\} \mathrm{d}\mathbf{q}},
    \\
    \left[ \mathbf{F}_{\mathbf{r}\tilde{\boldsymbol{\alpha}}} \right] _{m,n} &=\frac{2}{\sigma ^2}\int_{\mathcal{S} ^{(\mathrm{rx)}}}{\Re \left\{ \frac{\partial {E}\left( \mathbf{q} \right)}{\partial \left[ \mathbf{r} \right] _m}\frac{\partial {E}^*\left( \mathbf{q} \right)}{\partial \left[ \tilde{\boldsymbol{\alpha}} \right] _n} \right\} \mathrm{d}\mathbf{q}},
    \\
    \left[ \mathbf{F}_{\tilde{\boldsymbol{\alpha}}\tilde{\boldsymbol{\alpha}}} \right] _{m,n}& =\frac{2}{\sigma ^2}\int_{\mathcal{S} ^{(\mathrm{rx)}}}{\Re \left\{ \frac{\partial {E}\left( \mathbf{q} \right)}{\partial \left[ \tilde{\boldsymbol{\alpha}} \right] _m}\frac{\partial {E}^*\left( \mathbf{q} \right)}{\partial \left[ \tilde{\boldsymbol{\alpha}} \right] _n} \right\} \mathrm{d}\mathbf{q}},
\end{align}
where the expressions and derivations for $\mathbf{F}_{\mathbf{rr}}$, $\mathbf{F}_{\mathbf{r}\tilde{\boldsymbol{\alpha}}}$, and $\mathbf{F}_{\mathbf{r}\tilde{\boldsymbol{\alpha}}}^{}$ can be found in Appendix \ref{appendix:2}.
Following the above, the CRB matrix for estimating $\mathbf{r}$ can be calculated according to 
\begin{align}
    \mathrm{CRB}\left( \mathbf{F}_{\mathbf{rr}} \right) =\left[ \mathbf{F}_{\mathbf{rr}}-\mathbf{F}_{\mathbf{r}\tilde{\boldsymbol{\alpha}}}\mathbf{F}_{\tilde{\boldsymbol{\alpha}}\tilde{\boldsymbol{\alpha}}}^{-1}\mathbf{F}_{\mathbf{r}\tilde{\boldsymbol{\alpha}}}^{\mathrm{T}} \right] ^{-1}. \label{eq:crb_conti}
\end{align}

\section{CRB Optimization} \label{sect:crb_opti}
According to the CRB in \eqref{eq:crb_conti}, a mathematically tractable lower bound for sensing is established.
Therefore, minimizing CRB under the unit-power constraint can elevate the sensing performance.
Therefore, in this section, we first formulate the CRB minimization problem.
Then, we propose an SMGD method to solve this problem with reduced complexity.
Finally, a computational complexity analysis is attached.

\subsection{Problem Formulation}
Based on the derived CRB in \eqref{eq:crb_conti}, the optimization objective is to minimize CRB by designing a continuous source current function.
By doing so, the sensing performance can be enhanced \cite{li2008range}.
Here, we choose to minimize the trace of CRB, consistent with the approaches adopted for SPDA-enabled passive sensing scenarios, such as \cite{li2008range}, \cite{song2023intelligent}, and \cite{liu2022cramer}.
Specifically, jointly considering the transmit power budget over the Tx P-CAPA, the CRB optimization problem can be formulated as
\begin{subequations}
    \begin{align} 
        (\mathcal{P}_1)~&\min_{J\left( \mathbf{p} \right)} \,\,\mathrm{Tr}\left\{ \mathrm{CRB}\left( \mathbf{F}_{\mathbf{rr}} \right) \right\}  \notag \\ 
        &~{\rm{s.t.}}~ \int_{\mathcal{S} ^{(\mathrm{tx)}}}{\left| {J}\left( \mathbf{p} \right) \right|^2}\mathrm d\mathbf{p}\le P, \label{eq:P1-c1}
    \end{align}
\end{subequations}
in which the optimization objective is to minimize the trace of CRB, while \eqref{eq:P1-c1} indicates that the total power of the source current over the entire Tx P-CAPA cannot exceed the given power budget $P$.
Compared to conventional SPDAs \cite{song2023intelligent}, the crux of solving $(\mathcal{P}_1)$ is to design a continuous function $J\left( \mathbf{p} \right)$ instead of finding a vector or matrix of discrete variables.
Therefore, in the sequel, a lemma is first presented to identify the structure of the optimal solution and then followed by a Riemannian manifold algorithm to solve ($\mathcal{P}_1$).
This problem can be interpreted as designing the source current density $J(\mathbf p)$ for all $ \mathbf p \in \mathcal{S}^{\mathrm{(tx)}}$ to minimize the CRB at a position of interest, where an ST may be located. This interpretation is consistent with tracking systems, where the transmitter designs $J(\mathbf p)$ toward an estimated or predicted position to follow the target movement \cite{liu2022cramer}, thereby improving sensing accuracy around the optimized position. 
To assess robustness against initial localization errors, Fig. \ref{fig:9} shows that even with a coarse position estimate obtained from an unbiased estimator such as MLE, the optimization remains effective. Hence, our formulation aligns well with related works \cite{liu2022cramer,li2008range}. 
To further reduce the dependence of CRB on specific sensing parameters, one may adopt the Bayesian CRB (BCRB), which averages the CRB over the target distribution and enhances the FIM by incorporating prior information~\cite{scope2025bayesian}. 
Due to the absence of a closed-form expression for BCRB, we adopt CRB in this paper to provide insights into the sensing benefits of CAPA, while BCRB-based optimization is left for future work.

\subsection{Structure of Optimal Solution}
To make ($\mathcal{P}_1$) tractable, we prove that the optimal solution to ($\mathcal{P}_1$) lies within a subspace spanned by the transmit responses.
Therefore, the structure of the optimal solution is characterized by the following lemma:
\begin{figure*}[t!]
\begin{align}
    h_{1,n}&\triangleq \int_{\mathcal{S} ^{(\mathrm{tx)}}}{a_{\mathrm{t}}\left( \mathbf{k}_n \right) \mathbf{b}^{\mathrm{T}}\mathbf{w}\mathrm d\mathbf{p}}=\underset{\triangleq \mathbf{b}_{1,n}^{\mathrm{T}}\in \mathbb{C} ^{N\times 1}}{\underbrace{\left[ \int_{\mathcal{S} ^{(\mathrm{tx)}}}{a_{\mathrm{t}}\left( \mathbf{k}_n \right) \left[ \mathbf{b} \right] _1 \mathrm d\mathbf{p}},...,\int_{\mathcal{S} ^{(\mathrm{tx)}}}{a_{\mathrm{t}}\left( \mathbf{k}_n \right) \left[ \mathbf{b} \right] _N  \mathrm d\mathbf{p}} \right] }}\mathbf{w}^{}=\mathbf{b}_{1,n}^{\mathrm{T}}\mathbf{w}^{},\\
h_{2,n}^{\left( i \right)}&\triangleq \int_{\mathcal{S} ^{(\mathrm{tx)}}}{\left[ \nabla _{\mathbf{r}_n}a_{\mathrm{t}}\left( \mathbf{k}_n \right) \right] _i\mathbf{b}^{\mathrm{T}}\mathbf{w}\mathrm d\mathbf{p}}\notag \\
&=\underset{\triangleq \mathbf{b}_{2,n,i}^{\mathrm{T}}\in \mathbb{C} ^{N\times 1}}{\underbrace{\left[ \int_{\mathcal{S} ^{(\mathrm{tx)}}}{\left[ \nabla _{\mathbf{r}_n}a_{\mathrm{t}}\left( \mathbf{k}_n \right) \right] _i\left[ \mathbf{b} \right] _1 \mathrm d\mathbf{p}},...,\int_{\mathcal{S} ^{(\mathrm{tx)}}}{\left[ \nabla _{\mathbf{r}_n}a_{\mathrm{t}}\left( \mathbf{k}_n \right) \right] _i\left[ \mathbf{b} \right] _N \mathrm d\mathbf{p}} \right] }}\mathbf{w}=\mathbf{b}_{2,n,i}^{\mathrm{T}}\mathbf{w}, \quad i\in\{1,2,3\},\\
h_{3,n}^{\left( i \right)}&\triangleq c_0\alpha _n\,\,\left[ \nabla _{\mathbf{r}_n}a_{\mathrm{r}}\left( \boldsymbol{\kappa }_n \right) \right] _i, \quad i\in\{1,2,3\},
\\
h_{4,n}&\triangleq c_0\,a_{\mathrm{r}}\left( \boldsymbol{\kappa }_n \right) \alpha _n,
\end{align}
\hrulefill
\end{figure*}
\begin{lemma} \label{lemma:2}
    \normalfont The optimal source current function ${J}(\mathbf{p})$ for $(\mathcal{P}_1)$ can be expressed as a linear combination of transmit array responses, i.e., 
    \begin{align}
        J\left( \mathbf{p} \right) =\sum_{n=1}^N{w_ne^{\mathrm jk_0\left\| \mathbf{k}_n \right\|}}=\mathbf{b}^{\mathrm{T}}\mathbf{w},
    \end{align}
    where the weights vector and basis vector can be respectively expressed as
    \begin{align}
        \mathbf{w}&=\left[ w_1,w_2,...,w_N \right] ^{\mathrm{T}}\in \mathbb{C} ^{N\times 1},
        \\
        \mathbf{b}&=\left[ e^{j k_0\left\| \mathbf{k}_1 \right\| _2},...,e^{j k_0\left\| \mathbf{k}_N \right\| _2} \right] ^{\mathrm{T}}\in \mathbb{C} ^{N\times 1}.
    \end{align}
\end{lemma}
\begin{IEEEproof}
    Please refer to Appendix \ref{appendix:3}.
\end{IEEEproof}
Once the structure of the optimal solution is determined in light of \textbf{Lemma \ref{lemma:2}}, the rest is to determine the complex coefficient of each basis function, i.e., transmit array response.
To do so, we extract the complex coefficient vector $\mathbf{w}$ from the objective function and constraint of ($\mathcal{P}_1$).
As a preliminary step, the unit-power constraint presented by \eqref{eq:P1-c1} can be compactly expressed in a quadrature term, which is given by 
\begin{align}
    \int_{\mathcal{S} ^{(\mathrm{tx)}}}{\left| {J}\left( \mathbf{p} \right) \right|^2}d\mathbf{p}&=\int_{\mathcal{S} ^{(\mathrm{tx)}}}{\left| \mathbf{b}^{\mathrm{T}}\mathbf{w} \right|^2}\mathrm d\mathbf{p} \notag
    \\
    &=\int_{\mathcal{S} ^{(\mathrm{tx)}}}{\left( \mathbf{w}^{\mathrm{T}}\mathbf{b} \right) ^*\mathbf{b}^{\mathrm{T}}\mathbf{w}}\mathrm d\mathbf{p} \notag
    \\
    &=\mathbf{w}^{\mathrm{H}}\left( \int_{\mathcal{S} ^{(\mathrm{tx)}}}{\mathbf{b}^*\mathbf{b}^{\mathrm{T}}}\mathrm d\mathbf{p} \right) \mathbf{w} \notag
    \\
    &=\mathbf{w}^{\mathrm{H}}\mathbf{B}_0\mathbf{w} \le P,
\end{align}
where the entry of the cross integral matrix $\mathbf{B}_0$ can be expressed as
\begin{align}
    \left[ \mathbf{B}_0 \right] _{m,n}=\int_{\mathcal{S} ^{(\mathrm{tx)}}}{e^ { \mathrm j k_0\left\| \mathbf{k}_m \right\| } e^{ - \mathrm j k_0\left\| \mathbf{k}_n \right\| }}\mathrm d\mathbf{p}.
\end{align}

In the sequel, we will analyze the expression of the objective function of $(\mathcal{P}_1)$.
Before proceeding, we first define several terms for the clarity of presentation, which are shown at the top of the next page.
Given $n=\{1,...,N\}$, $i=\{1,2, 3\}$, and $j=\{1,2\}$, the partial derivatives in Appendix \ref{appendix:2} can be expressed as
\begin{align}
&\frac{\partial {E}\left( \mathbf{q} \right)}{\partial \left[ \boldsymbol{\xi } \right] _m}=\notag \\
&\begin{cases}
\overset{\mathbf{g}_{m}^{\mathrm{T}} \in \mathbb{C}^{N \times 1}}{\overbrace{\left( h_{3,n}^{(i)} \mathbf{b}_{1,n}^{\mathrm{T}} + h_{4,n} \mathbf{b}_{2,n,i}^{\mathrm{T}} \right)}} \mathbf{w},& \text{if } m = 3(n-1) + i \\
\underset{\bar{\mathbf{g}}_{m} \in \mathbb{C}^{N \times 1}}{\underbrace{\mathrm j^{i-1} \frac{h_{4,n}^{(i)}}{\alpha_n} \mathbf{b}_{1,n}}} \mathbf{w}^{\mathrm{T}}.& \text{if }m=3N+2(n-1)+j\\
\end{cases}
\end{align}
Therefore, we can further simplify the expression of the block matrix in the CRB expression as follows:
\begin{align}
    \left[ \mathbf{F}_{\mathbf{rr}} \right] _{m,n}&=\frac{2}{\sigma ^2}\int_{\mathcal{S} ^{(\mathrm{rx)}}}{\Re \left\{ \left( \mathbf{g}_{m}^{\mathrm{T}}\mathbf{w} \right) ^*\mathbf{g}_{m}^{\mathrm{T}}\mathbf{w} \right\} d\mathbf{q}}\notag \\
    &=\frac{2}{\sigma ^2}\Re \left\{ \int_{\mathcal{S} ^{(\mathrm{rx)}}}{\left( \mathbf{w}^{\mathrm{T}}\mathbf{g}_{m}^{} \right) ^*\mathbf{g}_{m}^{\mathrm{T}}\mathbf{w}d\mathbf{q}} \right\} \notag\\
    &=\frac{2}{\sigma ^2}\Re \left\{ \int_{\mathcal{S} ^{(\mathrm{rx)}}}{\mathbf{w}^{\mathrm{H}}\mathbf{g}_{m}^{*}\mathbf{g}_{m}^{\mathrm{T}}\mathbf{w}d\mathbf{q}} \right\} \notag\\
    &=\frac{2}{\sigma ^2}\Re \left\{ \mathbf{w}^{\mathrm{H}}\left[ \mathbf{B}_1 \right] _{m,n}\mathbf{w}^{} \right\} \notag \\
    &=\frac{2}{\sigma ^2}\Re \left\{ \mathrm{Tr}\left\{ \mathbf{w}^{}\mathbf{w}^{\mathrm{H}}\left[ \mathbf{B}_1 \right] _{m,n} \right\} \right\}  \notag \\
    &=\frac{2}{\sigma ^2}\Re \left\{ \mathrm{Tr}\left\{ \mathbf{W}\left[ \mathbf{B}_1 \right] _{m,n} \right\} \right\} ,
\end{align}
where $\mathbf{W}\triangleq \mathbf{w}^{}\mathbf{w}^{\mathrm{H}}$ satisfied $\mathrm{Rank}\left\{ \mathbf{W} \right\} =1$.
Thus, the entire $\mathbf{F}_{\mathbf{rr}}$ can be formulated as
\begin{align}
&\mathbf{F}_{\mathbf{rr}}= \notag \\
&\frac{2}{\sigma^2} \Re \left\{ 
    \left[ \begin{array}{ccc}
        \mathrm{Tr}\left\{ \mathbf{W} \left[ \mathbf{B}_1 \right]_{1,1} \right\} & \cdots & \mathrm{Tr}\left\{ \mathbf{W} \left[ \mathbf{B}_1 \right]_{3N,1} \right\} \\
        \vdots & \ddots & \vdots \\
        \mathrm{Tr}\left\{ \mathbf{W} \left[ \mathbf{B}_1 \right]_{3N,1} \right\} & \cdots & \mathrm{Tr}\left\{ \mathbf{W} \left[ \mathbf{B}_1 \right]_{3N,3N} \right\}
    \end{array} \right] 
\right\} \notag \\
&\triangleq \; \frac{2}{\sigma^2} \Re \left\{ \left( \mathbf{W} \otimes \mathbf{B}_1 \right)^{\mathrm{Tr}} \right\},
\end{align}
where $\mathbf{B}_1$ is a block matrix with entries $\left[ \mathbf{B}_1 \right] _{m,n}=\int_{\mathcal{S} ^{(\mathrm{rx)}}}{\mathbf{g}_{m}^{*}\mathbf{g}_{n}^{\mathrm{T}}\mathrm d\mathbf{q}}$ and $\mathbf{A}^{\mathrm{Tr}}$ denotes the blockwise trace operation of an arbitrary block matrix $\mathbf{A}$.
With a similar method, the other partial matrices can be expressed as
\begin{align}
    \mathbf{F}_{\boldsymbol{\alpha \alpha }}&=\frac{2}{\sigma ^2}\Re \left\{ \left( \mathbf{W}\otimes \mathbf{B}_2 \right) ^{\mathrm{Tr}} \right\} ,
    \\
    \mathbf{F}_{\mathbf{r}\boldsymbol{\alpha }}&=\frac{2}{\sigma ^2}\Re \left\{ \left( \mathbf{W}\otimes \mathbf{B}_3 \right) ^{\mathrm{Tr}} \right\} 
,
\end{align}
where the cross-integral matrices are given by
\begin{align}
     \left[ \mathbf{B}_2 \right] _{m,n}&=\int_{\mathcal{S} ^{(\mathrm{rx)}}}{\bar{\mathbf{g}}_{m}^{*}\bar{\mathbf{g}}_{n}^{\mathrm{T}}\mathrm d\mathbf{q}}, \\
     \left[ \mathbf{B}_3 \right] _{m,n}&=\int_{\mathcal{S} ^{(\mathrm{rx)}}}{\mathbf{g}_{m}^{*}\bar{\mathbf{g}}_{n}^{\mathrm{T}}\mathrm d\mathbf{q}}.
\end{align}
To calculate the integrals in CRB, we resort to the Gauss-Legendre numerical integral technique that can approximate an integral as a weighted summation. 
For simplicity, the implementation details of the Gauss-Legendre method are attached to Appendix \ref{appendix:4}.
Therefore, the optimization problem $(\mathcal{P}_1)$ can be recast as 
\begin{subequations}
    \begin{align}
    (\mathcal{P}_{1.1}) ~& \min_{\mathbf{W} \succeq 0}\,\,\frac{\sigma ^2}{2}\mathrm{Tr}\left\{ \Re \left\{ \left( \mathbf{W}\otimes \mathbf{B}_1 \right) ^{\mathrm{Tr}} \right\} -\Re \left\{ \left( \mathbf{W}\otimes \mathbf{B}_2 \right) ^{\mathrm{Tr}} \right\} \right. \notag\\
    &\left. \left( \Re \left\{ \left( \mathbf{W}\otimes \mathbf{B}_3 \right) ^{\mathrm{Tr}} \right\} \right) ^{-1} \notag \Re \left\{ \left( \mathbf{W}^{\mathrm{T}}\otimes \mathbf{B}_{2}^{\mathrm{T}} \right) ^{\mathrm{Tr}} \right\} \right\}
 \notag \\
        &\text{s.t.}~ \mathrm{Tr}\left\{ \mathbf{WB}_0 \right\} \le P, \label{eq:st1}\\
        &{\color{white}\text{s.t.}}~ \mathrm{Rank}\left\{ \mathbf{W} \right\} =1. \label{eq:st2}
    \end{align}
\end{subequations}
It is noted that the subspace method we proposed here is based on the exponential basis $e^{\mathrm{j}k_0\|\mathbf{k}_n\|}$ for $n\in \mathcal{N}$, which is the shared basis function of the round-trip channel responses at all STs and their respective derivatives.
As the size of CAPA is small compared with the propagation distances, we consider that the pathloss factors in channel responses and the derivatives are approximately invariant \cite{wang2023near, bjornson2021primer}.
Therefore, we treat them as constants that are absorbed into the optimization variables $\{ w_n \}_{n=1}^{N}$.
Therefore, when a particular polarization direction is considered, the basis function of channel response remains unchanged, thus indicating that our subspace method is still applicable in this case.

\subsection{Conventional Semi-Definite Relaxation Solution}
In problem $(\mathcal{P}_{1.1})$, we define $\mathbf{B}_1\left( \mathbf{W} \right) \triangleq \Re \{\left( \mathbf{W}\otimes \mathbf{B}_1 \right) ^{\mathrm{Tr}} \}$, $\mathbf{B}_2\left( \mathbf{W} \right) \triangleq \Re \{\left( \mathbf{W}\otimes \mathbf{B}_2 \right) ^{\mathrm{Tr}} \}
$, and $\mathbf{B}_3\left( \mathbf{W} \right) \triangleq \Re \{\left( \mathbf{W}\otimes \mathbf{B}_3 \right) ^{\mathrm{Tr}} \}$, respectively.
Based on the above definitions, we have 
\begin{align}
    \mathbf{K}\left( \mathbf{W} \right) =\mathbf{B}_1\left( \mathbf{W} \right) -\mathbf{B}_{2}^{}\left( \mathbf{W} \right) \mathbf{B}_{3}^{-1}\left( \mathbf{W} \right) \mathbf{B}_{2}^{\mathrm{T}} \left( \mathbf{W} \right).
\end{align}
Therefore, the original problem $(\mathcal{P}_{1.1})$ can be compactly written as
\begin{subequations}
    \begin{align}
    (\mathcal{P}_{1.2}) ~& \min_{\mathbf{W}\succeq 0} \quad \frac{\sigma^2}{2} \operatorname{Tr} \bigg\{   \mathbf{K}^{-1}\left( \mathbf{W} \right) \bigg\}
 \notag \\
        &\text{s.t.}~ \eqref{eq:st1}~\mathrm{and}~\eqref{eq:st2}. \notag
    \end{align}
\end{subequations}
To convert the above problem into a convex form, we utilize \textbf{Proposition 1} in \cite{wang2023stars}.
In particular, by introducing a auxiliary matrix $\mathbf{U} \succeq 0$ and the Schur complement, problem $(\mathcal{P}_{1.2})$ can be recast as 
\begin{subequations}
    \begin{align}
    (\mathcal{P}_{1.3}) ~& \min_{\mathbf{W}, \mathbf{U}\succeq 0} \quad \frac{\sigma^2}{2} \operatorname{Tr} \bigg\{   \mathbf{U}^{-1}) \bigg\} \label{obj:U}
 \notag \\
        &\text{s.t.}~ \eqref{eq:st1}~\mathrm{and}~\eqref{eq:st2} \notag \\
        &{\color{white}\text{s.t.}}~\mathbf{U} \succeq 0, \\
        &{\color{white}\text{s.t.}}~\left[ \begin{matrix}
	\mathbf{B}_1\left( \mathbf{W} \right) -\mathbf{U}&		\mathbf{B}_{2}^{}\left( \mathbf{W} \right)\\
	\mathbf{B}_{2}^{\mathrm{T}}\left( \mathbf{W} \right)&		\mathbf{B}_{3}^{}\left( \mathbf{W} \right)\\
\end{matrix} \right] \succeq 0 \label{eq:st3}
    \end{align}
\end{subequations}
Irrespective to the rank-1 constraint \eqref{eq:st2}, the objective function \eqref{obj:U} is a matrix decreasing on
the positive semidefinite matrix space, while constraint \eqref{eq:st3} ensures that $\mathbf{K}\left( \mathbf{W} \right) \succeq \mathbf{U}$.
As such, the equivalence between $(\mathcal{P}_{1.2})$ and $(\mathcal{P}_{1.3})$ can be guaranteed.
Furthermore, the only non-convex part is the rank-1 constraint.
To address this, we employ the semi-definite relaxation (SDR) to handle this constraint \cite{luo2010semidefinite}.
In particular, we first deliberately drop the non-convex constraint, i.e., \eqref{eq:st2}, yielding a convex optimization problem:
\begin{subequations}
    \begin{align}
    (\mathcal{P}_{1.4}) ~& \min_{\mathbf{W}, \mathbf{U}\succeq 0} \quad \frac{\sigma^2}{2} \operatorname{Tr} \bigg\{   \mathbf{U}^{-1} \bigg\}
 \notag \\
        &\text{s.t.}~ \eqref{eq:st1} \notag \\
        &{\color{white}\text{s.t.}}~\left[ \begin{matrix}
	\mathbf{B}_1\left( \mathbf{W} \right) -\mathbf{U}&		\mathbf{B}_{2}^{}\left( \mathbf{W} \right)\\
	\mathbf{B}_{2}^{\mathrm{T}}\left( \mathbf{W} \right)&		\mathbf{B}_{3}^{}\left( \mathbf{W} \right)\\
\end{matrix} \right] \succeq 0 \label{eq:st3}
    \end{align}
\end{subequations}
This problem can be solved optimally via CVX toolbox \cite{grant2014cvx}.
Then, we utilize the Gaussian randomization method to recover the rank-1 solution to the original problem.

\subsection{SMGD Algorithm}
To reduce the computational complexity in the large $N$ scenarios, we adopt the Riemannian manifold optimization method to solve $(\mathcal{P}_{1.1})$, whose superiority over the SDR method will be demonstrated in the analysis of the computational complexity and the simulation section.

\begin{algorithm}[t!]
\small
    \SetAlgoLined 
	\caption{Armijo Backtracking Line Search}\label{alg:1}
	\KwIn{Initial guess on step size $\upsilon_0$, the current Riemannian gradient $\eta$, the former search direction $\mathbf{d}$, maximum allowed iteration $T_{\max}$, control parameter $c=1\times10^{-4}$, objective function $F(\cdot)$, and shrinking parameter $\tau=0.5$.}
    \KwOut{Step size $\upsilon$.}
    Initialize the iteration counter $k=0$\;
    \For{$k \gets 0$ \KwTo $T_{\max}$}{
        \lIf{$F\left( \mathrm{retr}_{\mathbf{w}}\left\{ \mathbf{w}+\upsilon _k\mathbf{d} \right\} \right) \le F\left( \mathbf{w} \right) +c\upsilon _k\Re \left\{ \boldsymbol{\eta }^{\mathrm{H}}\mathbf{d} \right\}$}{break}
        \lElse{$\upsilon _{k}=\tau \upsilon _k$}
    }
    \lIf{$k=T_{\max}$}{$\upsilon=1\times 10^{-6}$}
    \lElse{$\upsilon=\upsilon _{k}$}
    \Return{The optimized step size $\upsilon$.}
\end{algorithm}
To adopt the SMGD method, we first release the rank-one constraint \eqref{eq:st2} by treating $\mathbf{W}$ as a vector multiplication, i.e., $\mathbf{w}\mathbf{w}^{\mathrm{H}}$.
Therefore, by optimizing $\mathbf{w} \in \mathbb{C}^{N \times 1}$, we can guarantee to obtain a rank-one solution.
Then, we deal with the power constraint \eqref{eq:st1}.
The optimal solution is obtained at the boundary of \eqref{eq:st1}. 
This can be proven using proof by contradiction, as a better solution can always be achieved by increasing the power until the boundary condition is met.
In this case, constraint \eqref{eq:st1} can be converted into a unit-power constraint represented by 
\begin{align}
    \mathbf{w}^{\mathrm{H}}\mathbf{B}_0\mathbf{w} = P. \label{eq:unit_pow_cons}
\end{align}
Given $\mathbf{B}_0 \in \mathbb{S}_{++}^N$ is a positive-defined matrix, \eqref{eq:unit_pow_cons} defines a complex ellipse manifold denoted by $\mathcal{M}$, on which the unit power constraint can be released.
For the simplicity of presentation, we denote the objective function of $(\mathcal{P}_{1.1})$ as $F(\mathbf{w})$.
Hence, the original optimization problem can be recast as
\begin{subequations}
    \begin{align}
    (\mathcal{P} _{2.2})~\mathop {\mathbf{w}=\mathrm{arg}\min _{\mathbf{w}\in \mathcal{M}}} \,\,F(\mathbf{w}),
    \end{align}
\end{subequations}
which is an unconstrained optimization on a manifold and can be solved efficiently by gradient descent.
To enable the gradient-descent-based method on the manifold $\mathcal{M}$, we must first define the gradients on the manifold $\mathcal{M}$, which represent the direction of the steepest increase of a function.
As a prerequisite step, the inner product between two vectors, $\left< \cdot, \cdot \right>$ on $\mathcal{M}$, can be evaluated as
\begin{align}
    \left< \mathbf{u},\mathbf{v} \right> =\mathbf{u}^{\mathrm{H}}\mathbf{B}_0\mathbf{v}.
\end{align}
where $\mathbf{u}$ and $\mathbf{v}$ are two points on the manifold $\mathcal{M}$.
Thus, $\mathcal{M}$ becomes a Riemannian manifold.
According to \cite{yu2016alternating} and \cite{alhujiali2019transmit}, each step of the gradient descent methods in the Euclidean space has its corresponding counterpart on the Riemannian manifold.
For a given point $\mathbf{w}$ on $\mathcal{M}$, the tangent space is defined by 
\begin{align}
    \mathcal{T} _{\mathbf{w}}\mathcal{M} \triangleq \left\{ \boldsymbol{\eta }\in \mathbb{C} ^{N\times 1}:\Re \left\{ \mathbf{w}^{\mathrm{H}}\mathbf{B}_0\boldsymbol{\eta } \right\} =0 \right\},
\end{align}
which describes a vector set perpendicular to $\mathbf{w}$.
Then, the Riemannian gradient defined by the steepest direction on $\mathcal{M}$ can be expressed by
\begin{align}
    \mathrm{grad}_{\mathbf{w}}\left\{ F\left( \mathbf{w} \right) \right\} &=\nabla _{\mathbf{w}}F\left( \mathbf{w} \right) -\notag \\&\qquad \qquad \frac{\mathbf{w}^{\mathrm{H}}\mathbf{B}_0\nabla _{\mathbf{w}}F\left( \mathbf{w} \right)}{\mathbf{w}^{\mathrm{H}}\mathbf{B}_0\mathbf{w}}\mathbf{B}_0\mathbf{w}=\boldsymbol{\eta }, \label{eq:gradient}
\end{align}
which can be geometrically interpreted as the maximum projection of the Euclidean gradient $\nabla _{\mathbf{w}}F\left( \mathbf{w} \right) $ to manifold $\mathcal{M}$.
The derivations of $\nabla _{\mathbf{w}}F\left( \mathbf{w} \right) $ can be obtained by applying a simplified chain rule for complex-valued variables in real-valued functions \cite{matrixcookbook}, which is omitted here for brevity.
Then, a search direction is defined by $\upsilon\mathbf{d}$, denoting a step of size $\upsilon>0$.
According to the Fletcher-Reeves (FR) method, the search direction can be specified by
\begin{align}
    \mathbf{d}^+=-\boldsymbol{\eta }^+ +\frac{\left\| \boldsymbol{\eta^+ } \right\| _{2}^{2}}{\left\| \boldsymbol{\eta }^- \right\| _{2}^{2}}\mathbf{d}^-, \label{eq:search_direction}
\end{align}
where the superscripts $+$ and $-$ denote the current and the previous iteration, respectively.
To enhance the algorithm's robustness, we utilize the Armijo backtracking line search algorithm summarized in \textbf{Algorithm \ref{alg:1}} to find the optimal $\upsilon$ at each iteration.
Furthermore, to ensure that the destination of $\upsilon\mathbf{d}$ is on the manifold $\mathcal{M}$, we define the retraction operator, that is characterized by
\begin{align}
    \mathrm{retr}_{\mathbf{w}}\left\{  \mathbf{w} + \upsilon\mathbf{d}\right\} \triangleq \frac{\mathbf{w} + \upsilon\mathbf{d}}{\sqrt{\Re \left\{ (\mathbf{w} + \upsilon\mathbf{d})^{\mathrm{H}}\mathbf{B}_0(\mathbf{w} + \upsilon\mathbf{d}) \right\}}}. \label{eq:retract}
\end{align}
Besides, to adopt the conjugated gradient method on $\mathcal{M}$, we must define a transport operation that maps two directions of two different tangent spaces.
This operation is distinguished by 
\begin{align}
    \mathrm{trans}_{\mathbf{w}^{-}\rightarrow \mathbf{w}^{+}}\left\{ \mathbf{d}^{-} \right\} = \mathbf{d}^{+}, \label{eq:transport}
\end{align}
where $\mathbf{w}^{+}$ and $\mathbf{w}^{-}$ are the point at the current and the former iteration, respectively.
Thus, the overall algorithm is summarized in \textbf{Algorithm \ref{alg:2}} \footnote{There are several stopping criteria for the manifold gradient descent method, such as the norm of two consecutive steps or the norm of the difference between successive gradients. The appropriate criterion can be determined through trial and error.}.
\begin{algorithm}[t!]
\small
    \SetAlgoLined 
	\caption{SMGD Algorithm for CRB Minimizing}\label{alg:2}
	\KwIn{Initial guess of weight vector $\mathbf{w}_0$, Rx/Tx CAPA coordinate sets $\mathcal{S}^{\mathrm{(rx)}}$ and $\mathcal{S}^{\mathrm{(tx)}}$, basis functions for STs, noise power $\sigma^2$, the number of points along the $x$- and $y$-axes $N_x$ and $N_y$, error-tolerant threshold $\delta$, maximum number of iterations $T_{\max}$.} 
	\KwOut{The optimized weight vector $\mathbf{w}$.}
  \tcp{Prerequisite Step:}\
  Calculate the initial gradient descent direction via $\mathbf{d}_0 = -\mathrm{grad}_{\mathbf{w}}\left\{F\left( \mathbf{w}_0 \right) \right\}$ according to \eqref{eq:gradient} and initialize $k=0$ \; 
 \tcp{Preforming Conjugate Gradient Descent:}\
 \While{$\left\| \mathbf{d}_{k+1}-\mathbf{d}_k \right\| _2\ge \delta$ and $k<T_{\max}$}{
Find the step size $\upsilon_k$ by \textbf{Algorithm \ref{alg:1}}\;
Find the next point $\mathbf{w}_{k+1}$ via retraction $\mathbf{w}_{k+1} = \mathrm{retr}_{\mathbf{w}}\left\{ \mathbf{w}_k+\upsilon _k\mathbf{d}_k \right\} 
$ according to \eqref{eq:retract}\;
Compute the Riemannian gradient at point $\mathbf{w}_{k+1}$ via $\boldsymbol{\eta }_{k+1}=\mathrm{grad}_{\mathbf{w}}\left\{ F\left( \mathbf{w}_{k+1} \right) \right\}$\;
Update the search direction via \eqref{eq:search_direction} and transport this direction via \eqref{eq:transport}\;
Step into the next iteration by $k=k+1$\;
}
\Return{The optimized weight vector $\mathbf{w} = \mathbf{w}_k$.}
\end{algorithm}
\begin{table*}[t!]
    \caption{System Parameters}
    \begin{center}
    \centering
    \resizebox{\textwidth}{!}{
        \begin{tabular}{|l|l|l||l|l|l|}
            \hline
            \centering
            $N$ & The number of sensing targets &$2$  & $f$ & The carrier frequency &$ 28~\mathrm{GHz}$\\
            \hline
            \centering
            $W^{\mathrm{(tx)}}_{\min}$ & The lower boundary of Tx CAPA's width &$-1$ m& $W^{\mathrm{(tx)}}_{\max}$ & The upper boundary of Tx CAPA's width &$0$ m\\
            \hline
            \centering
            $H^{\mathrm{(tx)}}_{\min}$ & The lower boundary of Tx CAPA's height &$-0.5$ m& $H^{\mathrm{(tx)}}_{\max}$ & The upper boundary of Tx CAPA's height &$0.5$ m\\
            \hline
            \centering
            \centering
            $W^{\mathrm{(rx)}}_{\min}$ & The lower boundary of Rx CAPA's width &$0$ m& $W^{\mathrm{(rx)}}_{\max}$ & The upper boundary of Rx CAPA's width &$1$ m\\
            \hline
            \centering
            $H^{\mathrm{(rx)}}_{\min}$ & The lower boundary of Rx CAPA's height &$-0.5$ m& $H^{\mathrm{(rx)}}_{\max}$ & The upper boundary of Rx CAPA's height &$0.5$ m\\
            \hline
            \centering
            $P$ & The transmit Power Tx CAPA &$100~\mathrm{mA}^2$& $\sigma^2$ & The noise power at Rx CAPA &$5.6 \times 10^{-3}~\mathrm{V^2/m^2}$ \\
            \hline
            \centering
            $\mathbf{r}_1$ & The position of the first ST &$[-5.0, 0.0, +5.0]$ m& $\mathbf{r}_2$ & The position of the second ST & $[+5.0, 0.0, +5.0]$ m \\
            \hline
            \centering
            $N_x/N_y$ & The number of GL points &$300$& $\alpha_n$ & The reflection coefficient of the $n$-th ST & $10+10 \mathrm j$ \\
            \hline
             \centering
            $\eta_0$ & The intrinsic impedance &$376.73~\Omega$&$T$ & Num. of Gaussian Randomization Trails & 100\\
            \hline
        \end{tabular}
    }
    \end{center}
    \label{tab:sim_params}
\end{table*}

\subsection{Analysis of Computation Complexity}
Given the number of targets is $N$, the computational complexity for SDR is specified by $\mathcal{O}(N^{3.5}/\epsilon_{\rm th})$ with $\epsilon_{\rm th}$ being the targeted solution accuracy and interior-point algorithm.
Then, the computational complexity for the Gaussian randomization process is given by $\mathcal{O}(TN^2)$ with $T$ being the number of trials for the Gaussian randomization process.
Therefore, the total computational complexity can be specified by $\mathcal{O}(N^{3.5}/\epsilon_{\rm th} + TN^2)$.
It is noted that, when the number of STs is small, the computational complexity of SDR is acceptable.
However, for large $N$, other methods, such as Manifold optimization techniques or machine learning-based methods, can be explored to achieve a lower complexity.
In the conventional wavenumber-domain sampling method, the number of optimization variables is given by $|\mathcal{K}_t| \approx \frac{\pi |\mathcal{S}^{(\mathrm{tx})}|}{\lambda^2}$, due to the spatial frequency sampling.
For high frequencies and planar arrays, the number of discrete samples becomes extremely large, resulting in prohibitive computational complexity. 
For example, when $f = 28~\mathrm{GHz}$ and $|\mathcal{S}^{(\mathrm{tx})}|=1~\mathrm{m}^2$, the required number of samples is approximately $2.74 \times 10^4$.
On the contrary, CAPA treats $J(\mathbf{p})$ as a continuous function, eliminating the need for explicit sampling..
Moreover, when combined with subspace methods, the computational complexity depends only on the number of STs rather than the discrete samples required in wavenumber-domain methods.

\section{Numerical Results} \label{section: simulation}

In this section, numerical results are presented to evaluate the performance of the proposed algorithm and to analyze the performance of the CAPA-based NISE systems.
The parameter settings are listed in Table \ref{tab:sim_params} and utilized throughout all the simulations unless otherwise specified.

\subsection{Convergence of the Proposed SMGD Algorithm}
\begin{figure}[t!]
    \centering
    \includegraphics[width=0.83\linewidth]{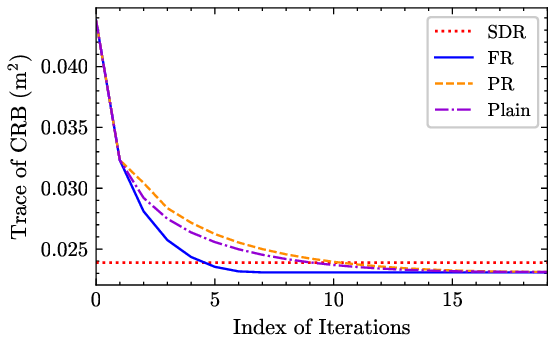}
    \caption{An illustration of the convergence behavior of SMGD algorithm over $100$ Monte Carlo simulations.}
    \label{fig:1}
\end{figure}
\begin{figure}[t!]
    \centering
    \includegraphics[width=0.93\linewidth]{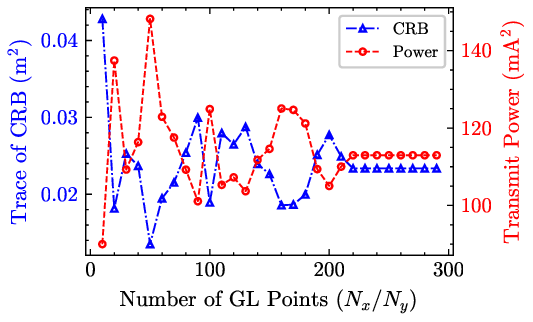}
    \caption{An illustration of the convergence behavior of the Gaussian-Legendre (GL) numerical integral.}
    \label{fig:2}
\end{figure}
\begin{figure}[t!]
    \centering
    \includegraphics[width=0.83\linewidth]{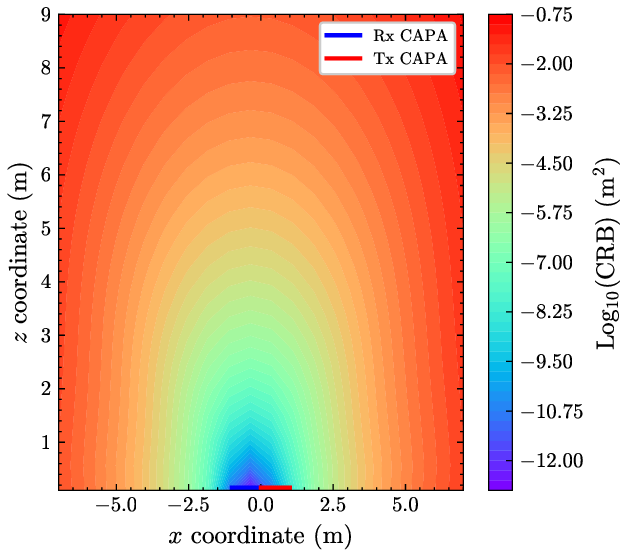}
    \caption{An illustration of achieved CRB over the $XOZ$ plane.}
    \label{fig:3}
\end{figure}
\begin{figure*}[htbp]
    \centering
    \subfloat[$\mathcal{S}^{(\mathrm{rx/tx})}=0.1~\mathrm{m}^2$\label{fig:4-1}]{
        \includegraphics[width=0.31\linewidth]{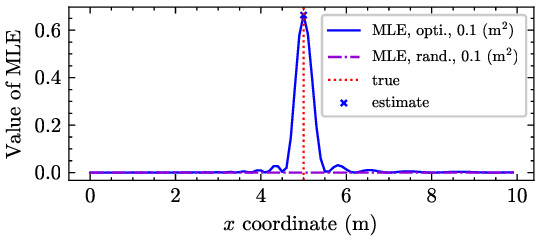}
    }
    \hfill
    \subfloat[$\mathcal{S}^{(\mathrm{rx/tx})}=0.5~\mathrm{m}^2$\label{fig:4-2}]{
        \includegraphics[width=0.31\linewidth]{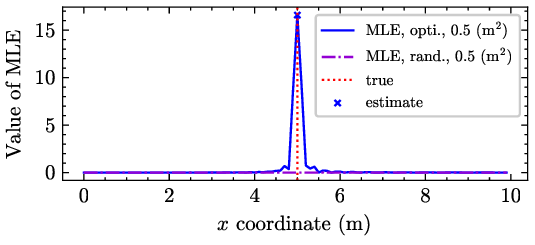}
    }
    \hfill
    \subfloat[$\mathcal{S}^{(\mathrm{rx/tx})}=1.0~\mathrm{m}^2$\label{fig:4-3}]{
        \includegraphics[width=0.31\linewidth]{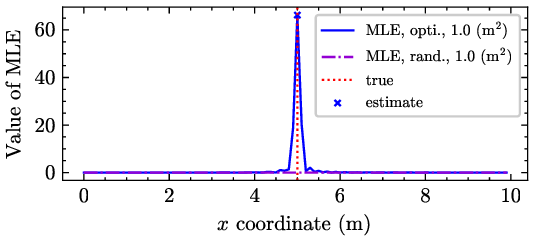}
    } \\
    \subfloat[$\mathcal{S}^{(\mathrm{rx/tx})}=0.1~\mathrm{m}^2$\label{fig:4-4}]{
        \includegraphics[width=0.31\linewidth]{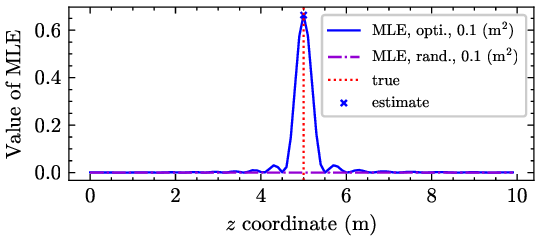}
    }
    \hfill
    \subfloat[$\mathcal{S}^{(\mathrm{rx/tx})}=0.5~\mathrm{m}^2$\label{fig:4-5}]{
        \includegraphics[width=0.31\linewidth]{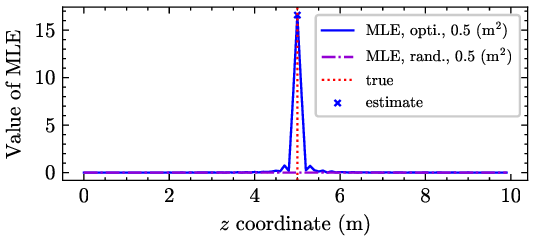}
    }
    \hfill
    \subfloat[$\mathcal{S}^{(\mathrm{rx/tx})}=1.0~\mathrm{m}^2$\label{fig:4-6}]{
        \includegraphics[width=0.31\linewidth]{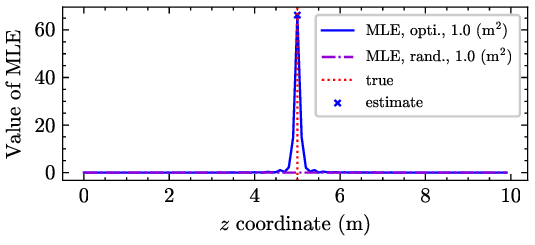}
    } 
    \caption{An illustration of the MLE spectrum along the $x$- and $z$-axes, respectively. ``opti." and ``rand." represent the proposed optimization method and the random policy, respectively. }
    \label{fig:6}
\end{figure*}
In Fig. \ref{fig:1}, we examine the convergence performance of the proposed SMGD method in comparison to multiple benchmarks: 1)~the SDR method with Gaussian randomization recovering, referred to as ``SDR", 2)~the plain conjugated gradient descent method on the manifold, referred to as ``Plain", and 3)~the Polak-Ribiere-based conjugated gradient descent method on the manifold, referred to as ``PR", respectively.
It is noted that the convergence behavior of ``SDR" is not shown here due to the large number of iterations needed for both the interior-point method and Gaussian randomization procedure for rank-$1$ recovery.
Instead, we present the final achieved throughput of ``SDR" for reference.
In addition, when the number of ST is small, ``SDR" can also be a good choice for CRB minimization.
Instead, the horizontal line of ``SDR" represents its averaged converged value, serving as a performance bound at the cost of high computational complexity.
As observed in Fig. \ref{fig:1}, both the ``FR" and ``PR" methods outperform the plain implementation of conjugated gradient descent, i.e., ``Plain".
In addition, ``FR" can converge more quickly compared to ``PR", thereby making it a more favorable choice for our problem.
More importantly, ``FR" can obtain a close performance to that of ``SDR" with a small number of iterations and reduced complexity.

In Fig. \ref{fig:2}, we present the convergence behavior of GL numerical integration by fixing the coefficient vector $\mathbf{w}$ as a random complex-valued vector that does not necessarily obey the unit-power constraint.
It can be seen that when $N_x/N_y$ exceeds $220$, the values of integral terms for both CRB and transmit power converge to fixed values, thereby justifying our setting for $N_x$ and $N_y$ in Table \ref{tab:sim_params}.
In contrast to the communication-only \cite{zhao2025continuous} and active sensing scenarios \cite{amico2022cramer}, the integrations require more GL points for convergence.
This can be attributed to the four-dimensional (4D) rather than two-dimensional (2D) integrations involved in CRB calculation.

\subsection{Evaluation of MLE and Beam Pattern}
Fig. \ref{fig:3} illustrates the achieved CRB over the $XOZ$ plane. 
Specifically, a single ST is placed on a grid with $400$ points over an area of $[-7~\mathrm{m}, +7~\mathrm{m}] \times [0.1~\mathrm{m}, 9~\mathrm{m}]$.
In addition, we utilize $\log_{10} \{\cdot \}$ function to post-process the achieved CRB.
The results reveal that: 1)~as ST moves closer to the Rx P-CAPA, a lower CRB is achieved, indicating an improved positioning accuracy, and 2)~the CRB distribution is symmetric with respect to the center point of the Rx P-CAPA.

Moreover, we present the MLE spectrum for the single ST scenarios, specifically for ``ST2" at $\mathbf{r}_2$ exists.
Using the derivations in Section \ref{sect:mle}, the MLE spectrum along $x$- and $z$-axes are depicted by  
Fig. \ref{fig:6}.
To provide a comparison, we also present a random policy, in which the source current is generated according to
\begin{align}
    {J}\left( \mathbf{p} \right) =\frac{1}{\sqrt{\left| \mathcal{S} ^{(\mathrm{tx)}} \right|}}\exp \left\{ \mathrm j \mathrm{Uniform}\left( -\pi ,+\pi \right) \right\}, \notag
\end{align} 
where $\mathrm{Uniform}\left( -\pi ,+\pi \right)$ can generate samples according to a uniform distribution in a range of $[ -\pi ,+\pi )$.
As demonstrated by Fig. \ref{fig:6}, the proposed method significantly outperforms the random policy, demonstrating its effectiveness in enhancing sensing accuracy.
From these sub-figures, it can be observed that, due to the dependence of the EM-based channel model on both $x$ and $z$ coordinates, CAPA-enabled NISE achieves a two-dimensional positioning, in contrast to the angle-only sensing in the conventional far-field scenarios.
More importantly, as the area of Tx and Rx P-CAPA increases, the spike in the MLE spectrum becomes sharper and taller with reduced side lobe amplitudes. 
This trend indicates that increasing the CAPA size enhances positioning accuracy.
 \begin{figure}[ht!]
        \centering
        \includegraphics[width=0.83\linewidth]{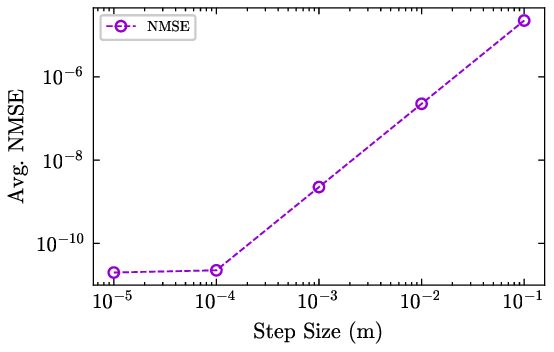}
        \caption{An illustration of NMSE as a function of step size.}
        \label{fig:nmse_vs_step_size}
    \end{figure}
Moreover, we present Fig. \ref{fig:nmse_vs_step_size} to show the normalized mean error (NMSE) versus the searching step size. The results in Fig. \ref{fig:2} indicate that more than 200 sampling points are required to achieve sufficient coverage, which can be attributed to the four-dimensional integration over both the transmit and receive planes. Consequently, even when the search interval is confined around the true target position, a finer search step still leads to thousands of evaluation points, resulting in prohibitive computational complexity. Therefore, we measure NMSE from two axes, respectively, rather than performing a two-dimensional search directly.
The simulation results are averaged over the two separate search dimensions. As shown in Fig. 10, when the step size is large, the accuracy is mainly determined by the step size, exhibiting a linear decay. In contrast, when the step size is sufficiently small, the accuracy is bounded by the inherent sensing resolution of the system, resulting in a flat performance curve. These results reveal that adopting a discrete search inevitably introduces a true-value-dependent bias in the large-step-size regime.

\begin{figure}[t!]
    \centering
    \subfloat[Beam Pattern for ST 1\label{fig:7-1}]{
        \includegraphics[width=0.5\linewidth]{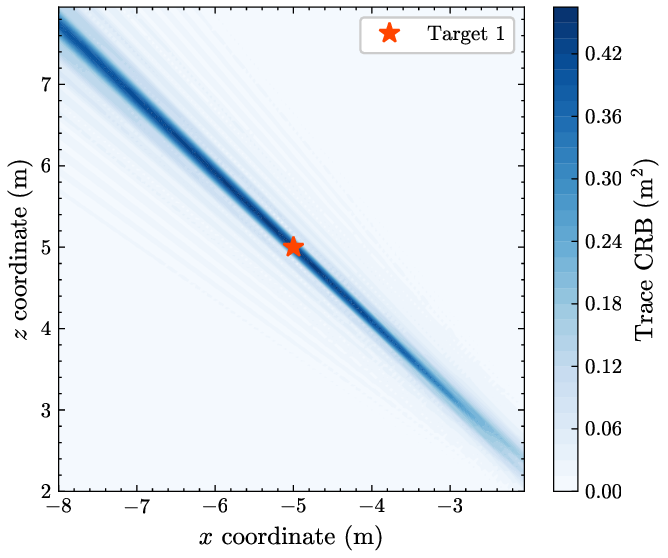}
    } 
    \subfloat[Beam Pattern for ST 2\label{fig:7-2}]{
        \includegraphics[width=0.5\linewidth]{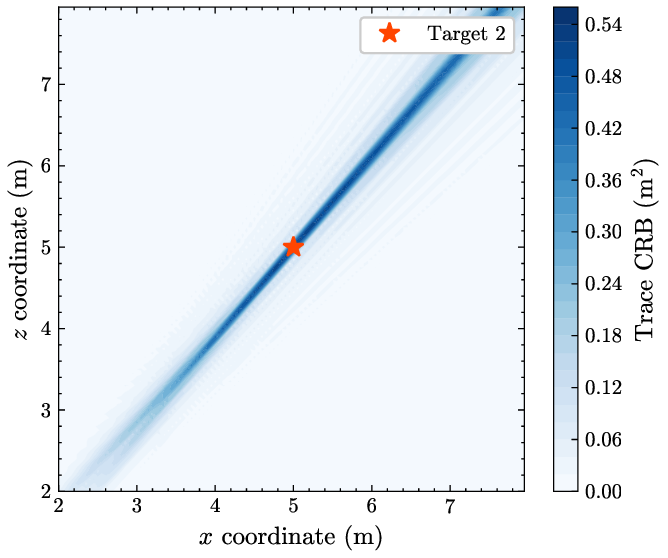}
    }
    \caption{An illustration of beam pattern for ST 1 at $[-5~\mathrm{m}, 0~\mathrm{m}, +5~\mathrm{m}]$ and ST 2 at $[+5~\mathrm{m}, 0~\mathrm{m}, +5~\mathrm{m}]$, respectively.}
    \label{fig:7}
\end{figure}

For the multi-ST scenario, the MLE calculation is high-dimensional, making it hard to visualize directly. 
As an alternative, we present the beam pattern, following the methodology outlined in  \cite{hua2024near}.
It is noteworthy that the path loss is eliminated in beam pattern calculations to remove the impact of distance on the results.
With these preliminaries, the beam patterns for ST 1 and ST 2 are depicted in Fig. \ref{fig:7}.
As shown in this figure, the energy of the beams is concentrated at the respective locations of ST 1 and ST 2, demonstrating the effectiveness of the proposed algorithm in the multi-ST scenario.

\subsection{Evaluation of CRB}
\subsubsection{CRB Versus Carrier Frequency}
\begin{figure}[t!]
    \centering
    \includegraphics[width=0.95\linewidth]{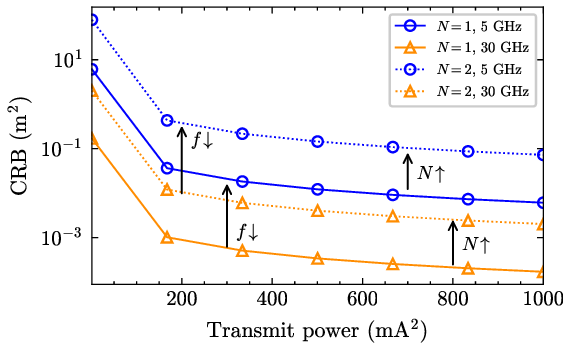}
    \caption{An illustration of the achieved CRB versus transmit power under different numbers of target $N$ and carrier frequency $f$.}
    \label{fig:8}
\end{figure}
In Fig. \ref{fig:8}, we analyze the achieved CRB as a function of transmit power under different numbers of the target $N$ and carrier frequency $f$.
From this figure, it can be observed that: 1)~As the transmit power increases, the CRB decreases correspondingly, aligning well with the conclusions drawn for the SPDA-based near-field sensing scenarios.
2)~As carrier frequency increases, a further reduction in CRB is observed, indicating an improved sensing accuracy;
3)~As more STs need to be positioned, the total CRB worsens, which can be attributed to the allocation of a limited total power budget, thus reducing the sensing accuracy for each target.

\subsubsection{Comparison Between CAPAs and SPDAs}
\begin{figure}[t!]
    \centering
    \includegraphics[width=0.95\linewidth]{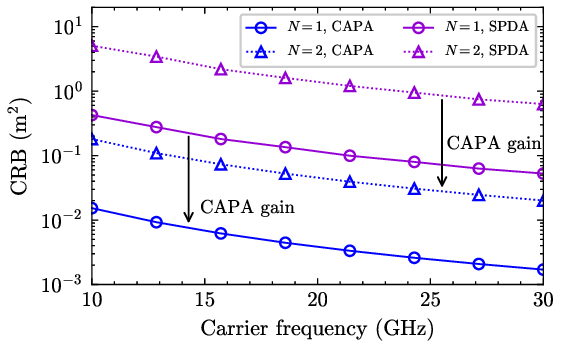}
    \caption{An illustration of the achieved CRB versus carrier frequency $f$ under different types of antenna architecture, i.e., CAPA and SPDA.}
    \label{fig:10}
\end{figure}
In Fig. \ref{fig:10}, we compare the investigated CAPA system to the conventional SPDA system to underscore the superiority of continuous aperture (CA).
The conventional SPDA system is analyzed by sampling the Rx and Tx antenna planes with $d=\frac{\lambda}{2}$ as spacing and $A_d = \frac{\lambda^2}{4\pi}$ as the effective aperture areas of each antenna.
The results demonstrate that the achieved CRB decreases as carrier frequency increases, thus reinforcing the results presented in Fig. \ref{fig:8}.
Moreover, compared to the conventional SPDA setup, the CAPA-based NISE achieves an order-of-magnitude improvement in accuracy, thereby underscoring the advantages of continuous aperture arrays.
Lastly, as the number of STs increases, the CRB also increases as depicted in Fig. \ref{fig:8}.

\subsubsection{CRB Versus Initial Estimation Error}
\begin{figure}[t!]
    \centering
    \includegraphics[width=0.95\linewidth]{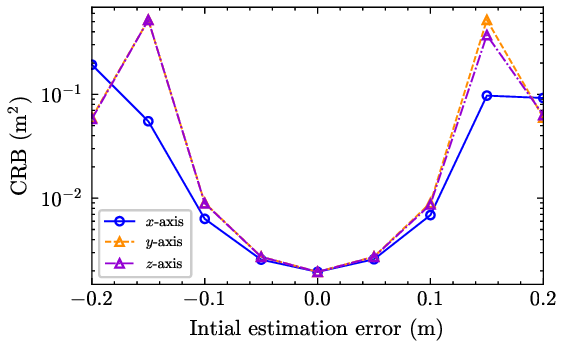}
    \caption{An illustration of the robustness of CRB optimization to initial position estimation results along $x$-, $y$-, $z$-axes for ST 1.}
    \label{fig:9}
\end{figure}
As highlighted in the pioneering work on CRB optimization \cite{li2008range}, CRB optimization requires prior knowledge of the target's position.
Therefore, it is essential to evaluate the robustness of the proposed algorithm to errors in the initial position estimation.
Fig. \ref{fig:9} illustrates the achieved CRB as a function of the initial estimation error. 
The results indicate that the CRB does not increase significantly when the error is within the range of $[-0.15~\mathrm{m}, -0.15~\mathrm{m}]$, demonstrating the robustness of the algorithm in this range.
It is important to note that since the target is located at $\mathbf{r}_1 = [-5.0~\mathrm{m}, 0.0~\mathrm{m}, 5.0~\mathrm{m}]$, and the Rx-CAPA center is at $[-5.0~\mathrm{m}, 0.0~\mathrm{m}, 5.0~\mathrm{m}]$, the error is symmetric only along the $y$-axis.
The asymmetry along the $x$- and $z$-directions arises from the asymmetric placement of the transmit and receive CAPAs, as their centers are not aligned on these axes.
\section{Conclusions} \label{sect:conclu}
This paper proposed a CRB minimization framework for CAPA-enabled NISE systems.
Specifically, different from existing research on active sensing with CAPA, the CRB expression was derived for a more practical passive sensing setup.
Based on the derivation, the CRB minimization problem was formulated, by solving which the sensing accuracy can be enhanced.
To make the optimization of the continuous source current function tractable, the structure of the optimal solution was proved, followed by the SMGD algorithm developed to obtain the coefficient vector with reduced computational complexity.
The numerical results verified the effectiveness of the proposed SMGD method and demonstrated the superiority of CAPAs over conventional SPDAs in terms of sensing performance.

\appendices
\section{The Derivations of CRB in \eqref{eq:crb_conti}}\label{appendix:2}
In light of \cite{stephen1993funda, li2008range, amico2022cramer}, the expression for the $(m, n)$-th entry in $\mathbf{F}_{\mathbf{rr}}$ can be expressed as 
\begin{align}
    &\left[ \mathbf{F}_{\mathbf{rr}} \right] _{m,n}=\frac{2}{\sigma ^2}\int_{\mathcal{S} ^{(\mathrm{rx)}}}{\Re \left\{ \frac{\partial {E}\left( \mathbf{q} \right)}{\partial \left[ \mathbf{r} \right] _m}\frac{\partial {E}^*\left( \mathbf{q} \right)}{\partial \left[ \mathbf{r} \right] _n} \right\} \mathrm{d} \mathbf{q}} \notag \\
    &=\frac{2}{\sigma ^2}\int_{H_{\min}^{\left( \mathrm{rx} \right)}}^{H_{\max}^{\left( \mathrm{rx} \right)}}{\int_{W_{\min}^{\left( \mathrm{rx} \right)}}^{W_{\max}^{\left( \mathrm{rx} \right)}}{\Re \left\{ \frac{\partial {E}\left( x,y \right)}{\partial \left[ \mathbf{r} \right] _m}\frac{\partial {E}^*\left( x,y \right)}{\partial \left[ \mathbf{r} \right] _n} \right\} \mathrm{d}x\mathrm{d}y}}. \label{eq:J_rr} \tag{A-1}
\end{align}
It is noted that, since both the Tx and Rx CAPAs are placed on the $XOY$ plane, we define ${E}\left( x,y \right) \triangleq \left. {E}\left( \mathbf{q} \right) \right|_{z=0}$ in \eqref{eq:J_rr}.
Then, based on \eqref{eq:conti_ele_field} and \eqref{eq:round-trip_channel}, we can factor out the summation term through
\begin{align}
    {E}(\mathbf{q})&\triangleq \int_{\mathcal{S} ^{(\mathrm{tx)}}}{h\left( \mathbf{q},\mathbf{p} \right) {J}(\mathbf{p})\mathrm{d}}\mathbf{p} \notag \\
    &=c_0\int_{\mathcal{S} ^{(\mathrm{tx)}}} {\sum_{n=1}^N{\,\,a_{\mathrm{r}}\left( \boldsymbol{\kappa }_n \right) \alpha _n}a_{\mathrm{t}}\left( \mathbf{k}_n \right) {J}(\mathbf{p})\mathrm{d}}\mathbf{p} \notag \\
    &\overset{\mathrm{(a)}}{=}c_0\sum_{n=1}^N{\,\,\int_{\mathcal{S} ^{(\mathrm{tx)}}}{a_{\mathrm{r}}\left( \boldsymbol{\kappa }_n \right) \alpha _n}}a_{\mathrm{t}}\left( \mathbf{k}_n \right) {J}(\mathbf{p})\mathrm{d}\mathbf{p} \notag \\
    &\overset{(\mathrm{b)}}{=} c_0\sum_{n=1}^N{\,\,a_{\mathrm{r}}\left( \boldsymbol{\kappa }_n \right) \alpha _n\int_{\mathcal{S} ^{(\mathrm{tx)}}}{a_{\mathrm{t}}\left( \mathbf{k}_n \right) {J}(\mathbf{p})\mathrm{d}\mathbf{p}}}, \notag
\end{align}
where $\mathrm{(a)}$ is obtained by leveraging the linear property of integration, and $\mathrm{(b)}$ is derived by extracting the terms independent of $\mathbf{p}$.
Following this, the partial derivative of ${E}(\mathbf{q})$ with respect to $\mathbf{r}$ can be computed.
In particular, for the $n$-th ST where $n \in \mathcal{N}$, the partial derivative with respect to $\mathbf{r}_n$ can be calculated via
\begin{align}
    \nabla _{\mathbf{r}_n}{E}\left( \mathbf{q} \right) &=\nabla _{\mathbf{r}_n} \left\{c_0\,\,a_{\mathrm{r}}\left( \boldsymbol{\kappa }_n \right) \alpha _n\int_{\mathcal{S} ^{(\mathrm{tx)}}}{a_{\mathrm{t}}\left( \mathbf{k}_n \right) {J}(\mathbf{p})\mathrm{d}\mathbf{p}} \right\} \notag
\\
&=c_0\alpha _n\,\,\nabla _{\mathbf{r}_n}a_{\mathrm{r}}\left( \boldsymbol{\kappa }_n \right) \int_{\mathcal{S} ^{(\mathrm{tx)}}}{a_{\mathrm{t}}\left( \mathbf{k}_n \right) {J}(\mathbf{p})\mathrm{d}\mathbf{p}} \notag \\
&\qquad +c_0\,\,a_{\mathrm{r}}\left( \boldsymbol{\kappa }_n \right) \alpha _n\int_{\mathcal{S} ^{(\mathrm{tx)}}}{\nabla _{\mathbf{r}_n}a_{\mathrm{t}}\left( \mathbf{k}_n \right) {J}(\mathbf{p})\mathrm{d}\mathbf{p}}, \label{eq:partial_r_n} \tag{A-2}
\end{align}
in which $\nabla _{\mathbf{r}_n}a_{\mathrm{t}}\left( \mathbf{k}_n \right) $ and $\nabla _{\mathbf{r}_n}a_{\mathrm{r}}\left( \boldsymbol{\kappa }_n \right)$ can be derived as
\begin{align}
\nabla _{\mathbf{r}_n}a_{\mathrm{t}}\left( \mathbf{k}_n \right) &=\nabla _{\mathbf{r}_n} \left\{ \frac{1}{\left\| \mathbf{k}_n \right\| _2}e^{ -\mathrm j k_0\left\| \mathbf{k}_n \right\| _2} \right\}
 \notag \\ 
&=-\frac{\mathbf{r}_n}{\left\| \mathbf{k}_n \right\| _{2}^{3}}\left( 1+\mathrm j k_0\left\| \mathbf{k}_n \right\| _2 \right) e^{ -\mathrm j k_0\left\| \mathbf{k}_n \right\| _2} , \tag{A-3} \label{eq:partial_a_t}\\
\nabla _{\mathbf{r}_n}a_{\mathrm{r}}\left( \boldsymbol{\kappa }_n \right) &=\nabla _{\mathbf{r}_n} \left\{  \frac{1}{\left\| \boldsymbol{\kappa }_n \right\| _2}e^{\mathrm j k_0\left\| \boldsymbol{\kappa }_n \right\| _2 } \right\}
 \notag \\ &=\frac{\mathbf{r}_n}{\left\| \boldsymbol{\kappa }_n \right\| _{2}^{3}}\left( 1- \mathrm j k_0\left\| \boldsymbol{\kappa }_n \right\| _2 \right) e^{\mathrm j k_0\left\| \boldsymbol{\kappa }_n \right\| _2}. \tag{A-4} \label{eq:partial_a_r} 
\end{align}
Similarly, the $(m, n)$-th entry in $\mathbf{F}_{\boldsymbol{\alpha \alpha}}$ we have 
\begin{align}
    \left[ \mathbf{F}_{\boldsymbol{\alpha \alpha }} \right] _{m,n}=\int_{H_{\min}^{\left( \mathrm{rx} \right)}}^{H_{\max}^{\left( \mathrm{rx} \right)}}{\int_{W_{\min}^{\left( \mathrm{rx} \right)}}^{W_{\max}^{\left( \mathrm{rx} \right)}}{\Re \left\{ \frac{\partial {E}\left( x,y \right)}{\partial \left[ \boldsymbol{\alpha } \right] _m}\frac{\partial {E}^*\left( x,y \right)}{\partial \left[ \boldsymbol{\alpha } \right] _n} \right\} \mathrm dx \mathrm dy}}, \label{eq:J_aa}\tag{A-5}
\end{align}
where the integrands are specified by
\begin{align}
    \frac{\partial {E}\left( \mathbf{q} \right)}{\partial \Re \left\{ \alpha _n \right\}}&= c_0\,a_{\mathrm{r}}\left( \boldsymbol{\kappa }_n \right) \int_{\mathcal{S} ^{(\mathrm{tx)}}}{a_{\mathrm{t}}\left( \mathbf{k}_n \right) {J}(\mathbf{p})\mathrm d\mathbf{p}} , \tag{A-6}\label{eq:partial_alpha_real}\\
    \frac{\partial {E}\left( \mathbf{q} \right)}{\partial \Im \left\{ \alpha _n \right\}}&= \mathrm j {c}_0\,a_{\mathrm{r}}\left( \boldsymbol{\kappa }_n \right) \int_{\mathcal{S} ^{(\mathrm{tx)}}}{a_{\mathrm{t}}\left( \mathbf{k}_n \right) {J}(\mathbf{p})\mathrm d\mathbf{p}}.\tag{A-7}\label{eq:partial_alpha_imag}
\end{align}
Finally, the $(m,n)$-th entries in $\mathbf{F}_{\mathbf{r}\boldsymbol{\alpha }}$ is defined by
\begin{align}
    \left[ \mathbf{F}_{\mathbf{r}\boldsymbol{\alpha }} \right] _{m,n}=\int_{H_{\min}^{\left( \mathrm{rx} \right)}}^{H_{\max}^{\left( \mathrm{rx} \right)}}{\int_{W_{\min}^{\left( \mathrm{rx} \right)}}^{W_{\max}^{\left( \mathrm{rx} \right)}}{\Re \left\{ \frac{\partial {E}\left( x,y \right)}{\partial \left[ \mathbf{r} \right] _m}\frac{\partial {E}^*\left( x,y \right)}{\partial \left[ \boldsymbol{\alpha } \right] _n} \right\} \mathrm dx \mathrm dy}}, \label{eq:J_ra} \tag{A-8}
\end{align}
where the integrands can be found in \eqref{eq:partial_r_n}, \eqref{eq:partial_alpha_real}, and \eqref{eq:partial_alpha_imag}.

\section{Proof of Lemma \ref{lemma:2}} \label{appendix:3}
To prove this lemma, we need to illustrate all the optimal solutions are located in the subspace $V$ spanned a set of basis functions (or the transmit array responses) $\mathcal{B} = \left\{ e^{ j k_0\left\| \mathbf{k}_n \right\| _2 } \right\} _{n\in \mathcal{N}}$.
Inspired by \cite{guo2024deep}, we utilize proof by contradiction.

First, we assume that an optimal solution ${J}^{\mathrm{opt.}}(\mathbf{p})$ to $(\mathcal{P}_1)$ can be found outside the subspace $V$.
Therefore, ${J}^{\mathrm{opt.}}(\mathbf{p})$ can be expressed by a component that is parallel to $V$ and a component that is perpendicular to $V$, thereby yielding
\begin{align}
    {J}^{\mathrm{opt}.}(\mathbf{p})={J}_{\bot}(\mathbf{p})+{J}_{\parallel}(\mathbf{p}), \tag{B-1}
\end{align}
where the perpendicular component ${J}_{\bot}(\mathbf{p})$ satisfies 
\begin{align}
    &\langle e^{ j k_0\left\| \mathbf{k}_n \right\| _2 } ,{J}_{\bot}(\mathbf{p})\rangle \notag \\
    &=\int_{\mathcal{S} ^{(\mathrm{tx)}}}{e^{\mathrm{j} k_0\left\| \mathbf{k}_n \right\| _2 } {J}_{\bot}(\mathbf{p}) \mathrm d}\mathbf{p}=0.~\mathrm{for}~\forall n \in \mathcal{N},\qquad \label{eq:C-2}\tag{B-2}
\end{align}
Then, we calculate the trace of CRB when ${J}^{\mathrm{opt}.}(\mathbf{p})$ is utilized for transmission.
In light of \eqref{eq:partial_r_n}, \eqref{eq:partial_alpha_real}, and \eqref{eq:partial_alpha_imag}, the following terms related to ${J}^{\mathrm{opt}.}(\mathbf{p})$ can be simplified by 
\begin{align}
    \mathrm{I}_{1,n} \notag &=  \int_{\mathcal{S} ^{(\mathrm{tx)}}}{a_{\mathrm{t}}\left( \mathbf{k}_n \right) {J}^{\mathrm{opt}.}(\mathbf{\mathbf{p}})\mathrm d\mathbf{p}}\\
    &=\int_{\mathcal{S} ^{(\mathrm{tx)}}}{a_{\mathrm{t}}\left( \mathbf{k}_n \right) {J}_{\parallel}(\mathbf{p})\mathrm d\mathbf{p}},
    ~~~~~\mathrm{for}~n \in \mathcal{N}\notag \\
    \mathrm{I}_{2, n} &= \int_{\mathcal{S} ^{(\mathrm{tx)}}}{\nabla _{\mathbf{r}_n}a_{\mathrm{t}}\left( \mathbf{k}_n \right) {J}^{\mathrm{opt}.}(\mathbf{p})\mathrm d\mathbf{p}} \notag \\
    &=\int_{\mathcal{S} ^{(\mathrm{tx)}}}{\nabla _{\mathbf{r}_n}a_{\mathrm{t}}\left( \mathbf{k}_n \right) {J}_{\parallel}(\mathbf{p})\mathrm d\mathbf{p}},~\mathrm{for}~n \in \mathcal{N} \notag
\end{align}
where the orthogonality in \eqref{eq:C-2} is utilized to cancel out the terms containing ${J}_{\bot}(\mathbf{p})$.
It can be observed that the perpendicular component ${J}_{\bot}(\mathbf{p})$ has no contributions to the trace of CRB.
Therefore, we can construct a better solution by concentrating power on the parallel component ${J}_{\parallel}(\mathbf{p})$.
Jointly considering the power budget, the new solution is given by 
\begin{align}
    {J}^{\prime}(\mathbf{p})&=\sqrt{\frac{P}{\int_{\mathcal{S} ^{(\mathrm{tx)}}}{\left| {J}_{\parallel}(\mathbf{p}) \right|^2 \mathrm d}\mathbf{p}}}{J}_{\parallel}(\mathbf{p})\notag \\
    &=\sqrt{\frac{\int_{\mathcal{S} ^{(\mathrm{tx)}}}{\left| {J}^{\mathrm{opt}.}(\mathbf{p}) \right|^2 \mathrm d}\mathbf{p}}{\int_{\mathcal{S} ^{(\mathrm{tx)}}}{\left| {J}_{\parallel}(\mathbf{p}) \right|^2 \mathrm d}\mathbf{p}}}{J}_{\parallel}(\mathbf{p})=\sqrt{c_1}{J}_{\parallel}(\mathbf{p})\notag
\end{align}
By capitalizing on the orthogonality, we have the following derivations:
\begin{align}
    \int_{\mathcal{S} ^{(\mathrm{tx)}}}{\left| {J}^{\mathrm{opt}.}(\mathbf{p}) \right|^2 \mathrm d}\mathbf{p}&=
\int_{\mathcal{S} ^{(\mathrm{tx)}}}{\left| {J}_{\bot}(\mathbf{p})+{J}_{\parallel}(\mathbf{p}) \right|^2 \mathrm d}\mathbf{p}
\notag\\
&=\int_{\mathcal{S} ^{(\mathrm{tx)}}}{\left| {J}_{\bot}(\mathbf{p}) \right|^2d}\mathbf{p}+\int_{\mathcal{S} ^{(\mathrm{tx)}}}{\left| {J}_{\parallel}(\mathbf{p}) \right|^2 \mathrm d}\mathbf{p}
\notag \\
&> \int_{\mathcal{S} ^{(\mathrm{tx)}}}{\left| {J}_{\parallel}(\mathbf{p}) \right|^2 \mathrm d}\mathbf{p}. \label{eq:C-3}\tag{B-3}\notag
\end{align}
According to \eqref{eq:C-3}, we have $c_1>1$.
Then, by substituting ${J}_{\parallel}(\mathbf{p})$ with ${J}^{\prime}(\mathbf{p})$, we have 
\begin{align}
\mathrm{I}_{1,n}^{\prime}=\sqrt{c_1}\mathrm{I}_{1,n}^{},~\mathrm{I}_{2,n}^{\prime}=\sqrt{c_1}\mathrm{I}_{2,n}^{},~\mathrm{for}~\forall n \in \mathcal{N}. \notag
\end{align}
Due to $\sqrt{c_1}$ being a constant number and irrelevant to the outer integral over $\mathcal{S}^{(\mathrm{rx})}$, we therefore have $\mathbf{F}_{\mathbf{rr}}^{\prime}=c_{1}^{}\mathbf{F}_{\mathbf{rr}}$, $\mathbf{F}_{\mathbf{r}\tilde{\boldsymbol{\alpha}}}^{\prime}=c_{1}^{}\mathbf{F}_{\mathbf{r}\tilde{\boldsymbol{\alpha}}}$, and $\mathbf{F}_{\tilde{\boldsymbol{\alpha}}\tilde{\boldsymbol{\alpha}}}^{\prime}=c_{1}^{}\mathbf{F}_{\tilde{\boldsymbol{\alpha}}\tilde{\boldsymbol{\alpha}}}$, where $\mathbf{F}_{\mathbf{rr}}$, $\mathbf{F}_{\mathbf{r}\tilde{\boldsymbol{\alpha}}}$, and $\mathbf{F}_{\tilde{\boldsymbol{\alpha}}\tilde{\boldsymbol{\alpha}}}$ are the partial FIM matrices parameterized by ${F}_{\parallel}(\mathbf{p})$ (or equivalently ${F}^{\mathrm{opt.}}(\mathbf{p})$).
Hence, we have the following inequality:
\begin{align}
    \mathbf{F}_{\mathbf{rr}}^{\prime}-\mathbf{F}_{\mathbf{r}\tilde{\boldsymbol{\alpha}}}^{\prime}\left( \mathbf{F}_{\tilde{\boldsymbol{\alpha}}\tilde{\boldsymbol{\alpha}}}^{\prime} \right) ^{-1}\left( \mathbf{F}_{\mathbf{r}\tilde{\boldsymbol{\alpha}}}^{\prime} \right) ^{\mathrm{T}}&=c_{1}^{}\left( \mathbf{F}_{\mathbf{rr}}-\mathbf{F}_{\mathbf{r}\tilde{\boldsymbol{\alpha}}}\mathbf{F}_{\tilde{\boldsymbol{\alpha}}\tilde{\boldsymbol{\alpha}}}^{-1}\mathbf{F}_{\mathbf{r}\tilde{\boldsymbol{\alpha}}}^{\mathrm{T}} \right) \notag \\
    &>\mathbf{F}_{\mathbf{rr}}-\mathbf{F}_{\mathbf{r}\tilde{\boldsymbol{\alpha}}}\mathbf{F}_{\tilde{\boldsymbol{\alpha}}\tilde{\boldsymbol{\alpha}}}^{-1}\mathbf{F}_{\mathbf{r}\tilde{\boldsymbol{\alpha}}}^{\mathrm{T}}. \notag
\end{align}
Since $\mathrm{Tr}\{(\cdot)^{-1}\}$ is a decreasing function over semi-definite matrix space, we have 
\begin{align}
    \mathrm{Tr}\left\{ \mathrm{CRB}\left( \mathbf{F}_{\mathbf{rr}}^{\prime} \right) \right\} <\mathrm{Tr}\left\{ \mathrm{CRB}\left( \mathbf{F}_{\mathbf{rr}}^{} \right) \right\}, \notag
\end{align}
which indicates that we find a better solution by constructing ${J}^\prime(\mathbf{p})$.
In this case, ${J}^{\mathrm{opt}.}(\mathbf{p})$ is not the optimal solution, which contradicts the initial assumption on ${J}^{\mathrm{opt}.}(\mathbf{p})$.
Consequently, the optimal solution only contains the component parallel to the subspace $V$.

\section{Implementation of Gauss-Legendre Numerical Integral}\label{appendix:4}
Letting the integrand be $f(x,y)$, we have the following derivations by performing a change of interval:
\begin{align}
    I&=\int_{\underline{H}}^{\overline{H}}{\int_{\underline{W}}^{\overline{W}}{f\left( x,y \right) \mathrm dx \mathrm dy}}
    \notag \\
    &=\int_{-1}^1{\int_{-1}^1{f\left( c_xx^{\prime}+b_x,c_y y^{\prime}+b_y \right) c_xc_y \mathrm dx^{\prime} \mathrm dy^{\prime}}}, \tag{C-1}\label{eq:integral}
\end{align}
where $x=c_xx^{\prime}+b_x$ with $c_x\triangleq 0.5\left( \overline{W}-\underline{W} \right) \,\,$ and $b_x\triangleq 0.5\left( \overline{W}+\underline{W} \right)$, and $y=c_yy^{\prime}+b_y
$ with $c_y\triangleq 0.5\left( \overline{H}-\underline{H} \right) $ and $b_y\triangleq 0.5\left( \overline{H}+\underline{H} \right) $, respectively. 
Based on where the integral is carried out, i.e., either on the Tx or Rx CAPAs, we select the following parameters accordingly: $\overline{W}=\{ W_{\max}^{\left( \mathrm{rx} \right)},W_{\max}^{\left( \mathrm{tx} \right)} \} 
$, $\underline{W}=\{ W_{\min}^{\left( \mathrm{rx} \right)},W_{\min}^{\left( \mathrm{tx} \right)} \} $, 
$\overline{H}=\{ H_{\max}^{\left( \mathrm{rx} \right)},H_{\max}^{\left( \mathrm{tx} \right)} \}$, and $\underline{H}=\{ H_{\min}^{\left( \mathrm{rx} \right)},H_{\min}^{\left( \mathrm{tx} \right)} \} 
$.
These parameters define the boundaries of the integration region for both the transmitting and receiving CAPAs.
Then, using the Gauss-Legendre quadrature method, the integral term in \eqref{eq:integral} can be approximated by
\begin{align}
    \mathrm{I}\approx \sum_{i=1}^{N_x}{\sum_{j=1}^{N_y}{w_i}w_jf\left( x_i,y_j \right) c_xc_y}, \tag{C-2}\label{eq:approx_sum}
\end{align}
where $x_i$ and $y_j$ are the roots of Legendre polynomials $P_{N_x}(x)$ and $P_{N_y}(y)$.
Here, the weights can be specified by
\begin{align} 
    w_i &= \frac{2}{\left(1 - x_i^2\right) [P_{N_x}^\prime (x_i)]^2}, \notag\\
    w_j &= \frac{2}{\left(1 - y_j^2\right) [P_{N_y}^\prime(y_j)]^2}. \notag
\end{align}
where ${P}_{N}^\prime \left( \cdot \right)$ represents the the first-order derivative of $N$-th Legendre polynomials evaluated at $(\cdot)=\{x_i, y_i\}$.
By capitalizing on the Gauss-Legendre quadrature method, we can convert all the integrals into multiple summations.

\end{document}